\documentclass[1L2pt,preprint]{aastex}

\begin{document}

\title{A Star Formation Law for Dwarf Irregular Galaxies}

\author{Bruce G. Elmegreen}
\affil{IBM Research Division, T.J. Watson Research Center, 1101 Kitchawan Road,
Yorktown Heights, NY 10598, bge@us.ibm.com}
\author{Deidre A. Hunter} \affil{Lowell
Observatory, 1400 West Mars Hill Road, Flagstaff, Arizona 86001, dah@lowell.edu}

\begin{abstract}
The radial profiles of gas, stars, and far ultraviolet radiation in 20 dwarf
Irregular galaxies are converted to stability parameters and scale heights for a
test of the importance of two-dimensional (2D) instabilities in promoting star
formation. A detailed model of this instability involving gaseous and stellar fluids
with self-consistent thicknesses and energy dissipation on a perturbation crossing
time give the unstable growth rates. We find that all locations are effectively
stable to 2D perturbations, mostly because the disks are thick.  We then consider
the average volume densities in the midplanes, evaluated from the observed HI
surface densities and calculated scale heights. The radial profiles of the star
formation rates are equal to about 1\% of the HI surface densities divided by the
free fall times at the average midplane densities. This 1\% resembles the efficiency
per unit free fall time commonly found in other cases. There is a further variation
of this efficiency with radius in all of our galaxies, following the exponential
disk with a scale length equal to about twice the stellar mass scale length. This
additional variation is modeled by the molecular fraction in a diffuse medium using
radiative transfer solutions for galaxies with the observed dimensions and
properties of our sample. We conclude that star formation is activated by a
combination of three-dimensional gaseous gravitational processes and molecule
formation. Implications for outer disk structure and formation are discussed.
\end{abstract}


\section{Introduction}
\label{intro}

The outer parts of spiral galaxies \citep{kennicutt89} and most of dwarf irregular
galaxies \citep{hunter96,meurer96,vanzee97,hunter98} appear to be marginally or
wholly stable according to the common Toomre condition for a single fluid disk of
zero thickness. However, star formation still occurs there
\citep{ferguson98,lelievre00,thilker05,gil05,hunter11}, suggesting either that the
outer disks are unstable anyway, or that some other triggering mechanism is
operating to form stars. The most inclusive stability models consider both stars and
gas in a thick disk \citep{romeo92,rafikov01,elmegreen11}. \cite{leroy08} showed
that with the Rafikov 2-fluid stability criterion, the main parts of spiral and
irregular galaxies are marginally unstable. \cite{yang07} showed the same thing for
the main part of the LMC. The far-outer parts of galaxies and dwarfs are still a
problem, though: the gas mass surface density usually dominates the stellar mass
surface density, and both get exponentially low with increasing radius, driving up
the stability parameter $Q$ \citep[e.g.,][]{yim14}.

Another difference between the inner and outer parts of galaxies is the slope of the
relationship between the star formation rate (SFR) per unit area and the gas surface
density. In the inner parts of disks, this slope varies between unity for the
molecular gas \citep{wong02,leroy08,bigiel08} and 1.5 or so for the total gas
(Kennicutt et al. 2007; see reviews in Kennicutt \& Evans 2012 and Dobbs et al.
2013). In the outer parts it can be much steeper, more like 2 to 3
\citep{bigiel08,bigiel10}. The linear relationship in the inner part is somewhat
sensible, showing that more gas makes more stars with a nearly constant consumption
time at the characteristic density of CO emission \citep{krum09,lada13}. The linear
law also gives a reasonable SFR over cosmic time \citep{feldmann13}. There are
assumptions in deriving this law, however, such as the background subtraction for
diffuse CO \citep{shetty13,shetty14} and star formation tracers
\citep{liu11,rahman11}, the level of variations in the CO excitation
\citep{momose13}, and possible variations in the conversion from CO to H$_2$
\citep{boissier03}. The quadratic slope in the outer part may contain the same
dependence on molecules as the inner part, although molecules are a trace component
in the outer part \citep{schruba11}, but also involve variations in the molecular
fraction and thermal phase of the gas \citep{ostriker10,krumholz13}.

The outer parts are also where the disk gets thick, presumably because of the
exponential drop in surface density $\Sigma$ combined with a slower decrease in the
vertical velocity dispersion $\sigma$ \citep{dib06,vdk11}. Such a flare follows from
the equation for scale height in an isothermal gas, $H=\sigma^2/\left(\pi G
\Sigma\right)$. The disk flare means that the midplane density drops doubly fast,
because it equals the ratio of $\Sigma$ to $2H$. If the fundamental star formation
law is three-dimensional \citep[3D; e.g.,][]{ferguson98,elmegreen02,krumholz12}, and
not two-dimensional (2D) like the spiral wave instability originally targeted for
the Q parameter \citep{toomre64}, then the change from a linear to a quadratic star
formation law with radius could result in part from the flare
\citep[e.g.,][]{barnes12}.

Here we consider the applicability of the 2D, thick-disk gravitational instability
model to 20 dwarf irregular (dIrr) galaxies that have most of their star formation
in the quadratic regime \citep{bigiel08} and are among the class of galaxies that
have relatively thick disks \citep{hodge66,vdb88,staveley92}. Equilibrium models of
the disk thicknesses in four dwarf galaxies, based on HI and other data in the
THINGS survey, were also in \cite{banerjee11}. Our analysis of thickness is similar
to theirs, based on the multi-component model of vertical equilibrium in
\cite{narayan}.

We find that, considering disk thickness, nearly all of our dIrrs are effectively
stable in two dimensions throughout. That is, their instability growth times are all
very long, as calculated for a gas+star mixture with equilibrium disk thicknesses
and gaseous energy dissipation on a perturbation crossing time, as in a realistic
turbulent medium \citep{elmegreen11}. This appears to rule out 2D instabilities as a
model for star formation. We then investigate the possible role of 3D processes by
considering the dynamical time at the midplane density, rather than the growth time
of a 2D instability. We consider that the SFR per unit area scales with the product
of the HI gas column density and the dynamical rate from the 3D density \citep[as
in, e.g.,][]{elmegreen02,krum09,krumholz12}. The multiplicative factor that connects
these two rates turns out to be the usual $\sim 1$\% \citep{krumholz07}.  Thus, star
formation still follows 3D gravitational processes even where the conventional
Toomre Q parameter predicts a high level of 2D stability.

We also find a residual dependence of this multiplicative factor on radius, for
which two models are considered. The first model has the SFR connected to the HI
dynamical rate by an efficiency factor $\epsilon_{\rm ff}$ that is proportional to
the molecular fraction when both the molecular and atomic layers have the same
radial dependence for thickness. In this model, the radiation field that determines
the molecular fraction has a volume emissivity proportional to the stellar density.
The second model has the SFR connected directly to the H$_2$ dynamical rate with a
constant efficiency and assumes the thickness of the molecular layer is constant
with radius. For this, the radiation field that determines the molecular fraction
has a volume emissivity proportional to the density of FUV radiation. In both cases,
the molecular fraction is determined from radiative transfer solutions in a diffuse
medium using the observed or derived average gradients in stellar density and
intensity, gas density, and scale height.

Additional models consider the variation of the molecular fraction with radius using
the formulism in \cite{krum09,krumholz09b} and using another method that integrates
over the probability distribution function for cloud column density with a
separation between high column clouds that form molecules and low column clouds that
do not. In both cases, the usual threshold behavior for the formation of H$_2$
appears, suggesting a sudden drop in molecular fraction at some radius in the
galaxy. Such a drop is not observed in the star formation rates here and may not be
appropriate for the far-outer regions of galaxies which presumably have a low
molecular fraction throughout.

In what follows, Section \ref{obs} describes the observations, Section
\ref{scaleheights} demonstrates that the dIrrs in our survey have relatively thick
disks, Section \ref{two-fluid} shows they are significantly stabilized by thickness
against 2D disk instabilities, and Section \ref{threed} considers a fundamental star
formation law based on 3D dynamical processes with some efficiency. Section
\ref{radial} looks at the radial trends in our data for the purpose of understanding
a radial trend found in the efficiency. Section \ref{mdc} considers that this
efficiency trend results from variations in the molecular fraction in a diffuse
interstellar medium, which depends on molecular self-shielding and mutual shielding
inside of and between diffuse clouds.  The molecular fraction depends on the
radiation field, so we solve the equation of radiative transfer for a dIrr galaxy in
Section \ref{radtrans} using the average radial trends for gaseous and stellar
densities and scale heights. A model using FUV radiation for molecular dissociation
(Section \ref{secondsol}) requires a uniform line-of-sight thickness for the H$_2$
clouds, rather than a disk flare as inferred for the HI.  Other models considering a
threshold behavior for H$_2$ formation are in Section \ref{sharp}. A discussion
about the implications of these results for the outer edges of galaxies is in
Section \ref{search}. The conclusions are in Section \ref{conc}.

\section{Observations}
\label{obs}

The 20 dIrr galaxies used in this study are a sub-sample of the 41 galaxies that
constitute the LITTLE THINGS (Local Irregulars That Trace Luminosity Extremes, The
H$\,${\sc i} Nearby Galaxy Survey) sample \citep{hunter12}. The galaxies listed in
Table \ref{tab-sample} were chosen from the larger sample to be those having stellar
mass density profiles in \cite{zhang12} and H$\,${\sc i} rotation curves in
\cite{oh15}. The galaxies are relatively nearby ($<$10.3 Mpc), gas-rich, and have
recent star formation.  One galaxy, the Blue Compact Dwarf Haro 29, is suspected of
having undergone a recent interaction \citep{ashley13}. Otherwise, the sample
galaxies are expected to be representative of dIrrs with on-going star formation
through primarily internal processes.

The gas data used here were derived from cubes of H$\,${\sc i}  emission as a
function of velocity obtained with the Very Large Array (VLA\footnote{ The VLA is a
facility of the National Radio Astronomy Observatory (NRAO). The NRAO is a facility
of the National Science Foundation operated under cooperative agreement by
Associated Universities, Inc. These data were taken during the upgrade of the VLA to
the Expanded VLA, now JVLA.}). The observations, calibration and mapping procedures
are described by \cite{hunter12}. The rotation curves were determined using a
tilted-ring model through an iterative process that deconvolves bulk rotation from
non-circular motions (Oh et al.\ 2011, 2014). Here, we use the observed rotation
curves before an asymmetric drift correction is applied because that is appropriate
for the actual gas angular momentum and the value of $\kappa$ in the Toomre
parameter $Q$. The H$\,${\sc i} mass density profiles were determined from the
Naturally-weighted integrated H$\,${\sc i} (moment zero) maps using the parameters
determined from the rotation curve fitting (center, position angle, inclination,
annuli width). The H$\,${\sc i} mass was multiplied by 1.36 to include He.
Uncertainties in the azimuthally averaged gas mass are estimated to be $<10$\%. The
gas velocity dispersion profiles were measured from the flux-weighted dispersion
(moment 2) maps also derived from the Naturally-weighted H$\,${\sc i} cubes.

The stellar mass surface density profiles are from \cite{zhang12}. That study
performed spectral energy distribution (SED) fitting to surface photometry from {\it
Galaxy Evolution Explorer} \citep[{\it GALEX},][]{martin05} FUV and NUV, Lowell
Observatory $UBV$ and H$\alpha$, and {\it Spitzer} \citep{werner04} 3.6 $\mu$m
images \citep{fazio04}. The stellar azimuthally-averaged surface photometry used
ellipse parameters appropriate to the stellar disk and determined from the $V$-band
image \citep{hunter06}. Modeling the photometry used a library of $4 \times 10^6$
star formation histories and lifetimes divided into six logarithmic age bins.  A
Chabrier IMF was assumed.

The stellar velocity dispersions are evaluated from the expression $\sigma_{\rm
star}=10.0^{-0.15M_{\rm B}-1.27}$ km s$^{-1}$ \citep{swaters99,johnson12,johnson15}
for B-band absolute magnitude $M_{\rm B}$. These dispersion are assumed to be
constant with radius.

The current SFRs are taken as proportional to the FUV surface photometry, given in
\cite{hunter10} and updated by \cite{zhang12} to use a more recent data reduction
pipeline and to include all of the LITTLE THINGS galaxies. The surface brightnesses,
$\mu_{\rm FUV}$, used here are AB magnitudes per square arcsec corrected for
extinction. The FUV surface photometry is converted to a SFR surface density based
on a multiplicative factor between $L_{\rm FUV}$ and SFR$_{\rm FUV}$ given by
\cite{kennicutt98} and modified for the sub-solar metallicities of the dIrrs by
\cite{hunter10}. The result is $\Sigma_{\rm SFR}=10^{-0.4\mu_{\rm
FUV}+7.415}\;M_\odot$ pc$^{-2}$ Myr$^{-1}$ for a Salpeter IMF. For a Chabrier IMF,
the mass is smaller by the factor 0.55, and that is what we use here: $\Sigma_{\rm
SFR}=10^{-0.4\mu_{\rm FUV}+7.155}\;M_\odot$ pc$^{-2}$ Myr$^{-1}$. The FUV flux
density is obtained by the usual formula, $F_{\rm FUV}=10^{-0.4(\mu_{\rm
FUV}+48.6)}$ erg cm$^{-2}$ s$^{-1}$ Hz$^{-1}$ arcsec$^{-2}$.

The stellar mass densities, gas densities, rotation velocities, gas velocity
dispersions, and FUV surface brightnesses were collected from azimuthally-averaged
surface photometry and surface density profiles. The intersection of these data sets
is 20 galaxies.  The widths of the annuli used were fixed for each galaxy, but vary
from galaxy to galaxy: 20 pc (IC1613) to 340 pc (DDO 52), with a median over the 20
galaxies of 120 pc.  This is the spatial resolution of our study.

The uncertainties in the SED fittings that produce the stellar masses increase as
one goes to the outer disk. For example, in CVnIdwA, one of the fainter dwarfs, the
last annulus in which we quote a stellar mass density gives the value $\log(M_\odot
\; {\rm pc}^{-2}) = -0.88^{+0.24}_{-0.20}$. However, as the uncertainty in the
stellar mass climbs, the stellar mass density itself drops and the total mass is
dominated by the gas. Therefore, the contribution of the stellar mass uncertainty to
the total baryonic mass uncertainty is small. The uncertainty of the FUV surface
photometry also increases as the surface brightness drops. We stop when the
uncertainty reaches 0.5 mag.

The derivative of the rotation curve at a particular radius, which is used to
evaluate the epicyclic frequency $\kappa$, was obtained from the difference between
the rotation values on either side of that radius. Irregularities in the rotation
curves produce irregularities in $\kappa$, but they are usually not large. The local
values for the derived quantities do not influence the main conclusions here anyway.

\section{Two-dimensional gravitational stability at low surface density with a thick disk}
\label{scaleheights}

Two-dimensional stability in a disk can be measured by the Toomre parameter
$Q=\kappa\sigma/(3.36 G\Sigma)$ for epicyclic frequency $\kappa$, radial velocity
dispersion $\sigma$, and mass surface density $\Sigma$. The factor 3.36 is for a
stellar fluid \citep{toomre64}; it should be $\pi$ for a gaseous fluid
\citep{safronov60}. In a flat rotation curve, $\kappa\sim1/R$ for galactocentric
radius $R$, and for a typical exponential disk, $\Sigma\propto e^{-R/R_D}$, so that
$Q\propto e^{R/R_D}/R$ for nearly constant $\sigma$. Such a $Q$ is a decreasing and
then increasing function of radius. The increase in the outer part of the disk
implies these regions should be stable. More realistic analyses could consider
combined gas+star disks, disks with several components of gas at different
temperatures, disks with stars plus a highly dissipative turbulent gas, disks with
spiral waves that might trigger local gravitational collapse even with a high
average $Q$, etc.

Disk thickness also has an effect. It decreases the in-plane force of gravity per
unit mass column density in a disk because the part of the mass that is far from the
midplane directs the force vector away from the plane.  \cite{vandervoort70} showed
that in 2D models, the disk mass column density $\Sigma$ should be decreased by the
factor $(1+kH)^{-1}$ to account for this, where $k$ is the wavenumber (inverse
length) of a column density perturbation and $H$ is the disk half thickness. A more
detailed study of thickness effects in \cite{elmegreen11} confirms this factor and
shows that it is accurate to within 12\% for a wide range of conditions, compared to
a directly integrated force in three dimensions. Because dIrr galaxies and the outer
parts of spiral galaxies in general are thought to be relatively thick, there should
also be a trend toward greater stability in these regions compared to the
conventional 2D analysis just from thickness effects.

\begin{figure}\epsscale{.9}
\plotone{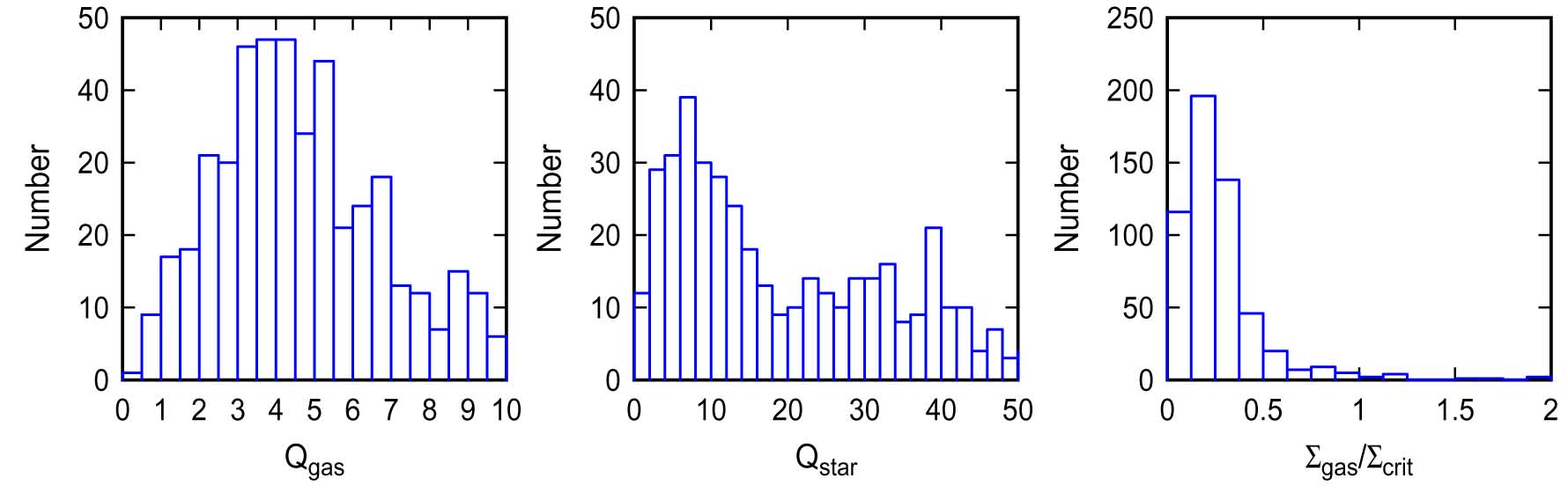}
\caption{(left and center) The distribution of gaseous and stellar $Q$ values for all of the radii and galaxies in our survey.
(right) The distribution of the ratio of the local HI surface density and the Kennicutt (1989)
critical density for star formation, based on the 2D stability analysis. The dIrr galaxies
considered here have high values of $Q_{\rm gas}$ and very high values of
$Q_{\rm star}$, making them effectively stable against radial motions and spiral waves driven by
self-gravity. The low values of $\Sigma_{\rm gas}/\Sigma_{\rm crit}$ suggest the
same thing. } \label{hunter_q_hqandscsg}\end{figure}

The radial profiles compiled for the present survey (Sect. 2) provide good
confirmation of these points. Figure \ref{hunter_q_hqandscsg} shows histograms of
the conventional 2D instability parameter $Q=\kappa\sigma/(\pi G \Sigma)$ for gas
alone and for stars alone (using $\pi$ in the denominator for stars too), measured
at each radius in each galaxy. $Q$ is generally much larger than unity, which
implies stability. Also shown is a histogram of the ratio of the
azimuthally-averaged gas surface density to the critical surface density
$\Sigma_{\rm crit}=0.7\kappa\sigma/(3.36G)$ based on the condition $Q=1.4$ from
\cite{kennicutt89}. This ratio is low, meaning the disks are stable in a
conventional sense, in agreement with the high values of $Q$ for the gas. The $Q$
values are so large that the 2-fluid disk should also be stable, if we consider that
the effective $Q$ value is the harmonic sum of the gaseous and stellar values
\citep{wang94}. Thickness corrections to this harmonic sum from \cite{romeo11} make
the disks even more stable in 2D.

\begin{figure}\epsscale{.9}
\plotone{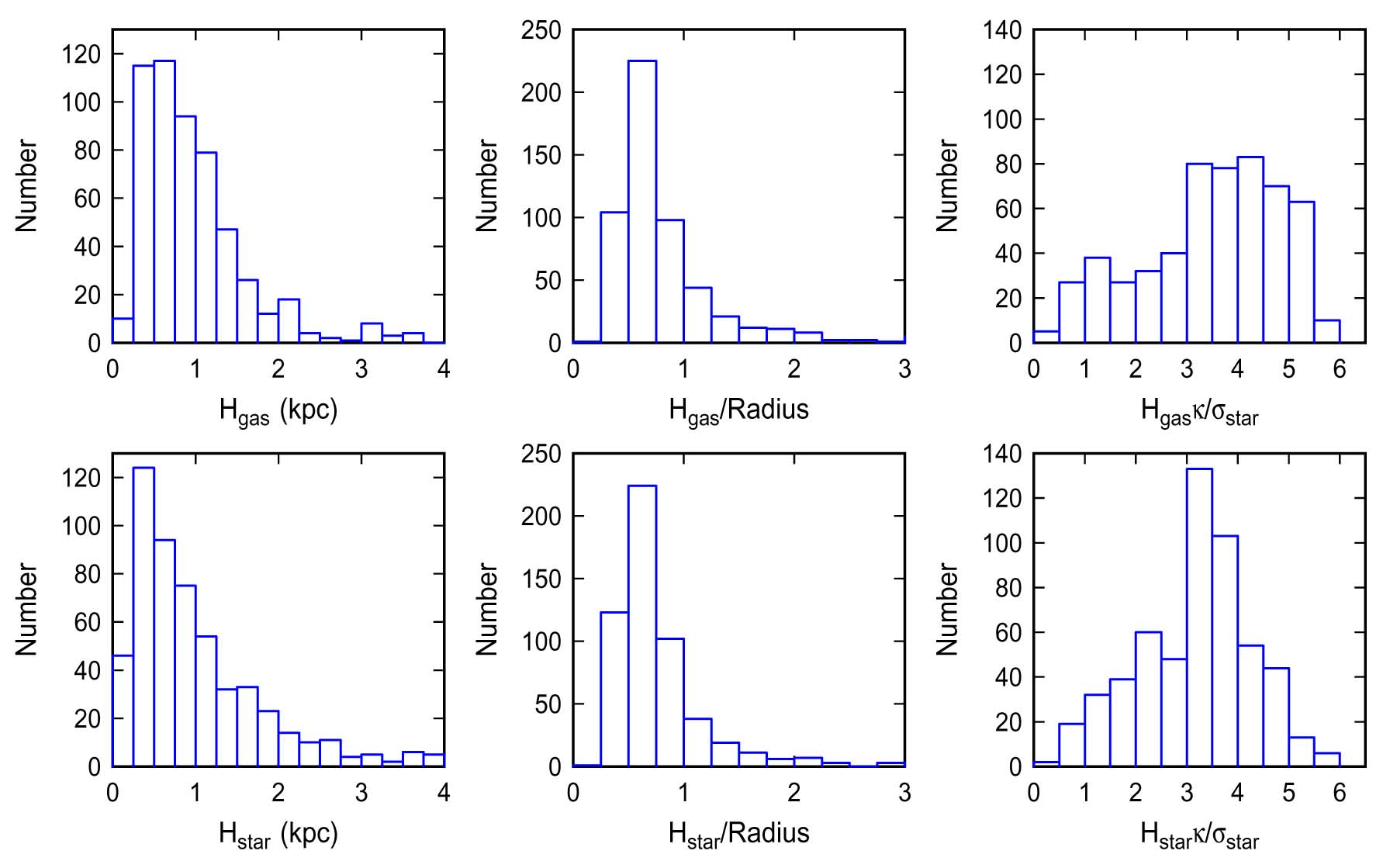}
\caption{The distributions of vertical scale heights for gas (top) and stars (bottom) for all
of the radii and galaxies in our survey.  The scale height is given in three forms: absolute (left),
relative to
the local radius (center), and relative to the local epicyclic scale (right).
By all measures, the scale heights
are large, which weakens self-gravitational forces parallel to the disks and makes the
2D gravitational instability even less effective than indicated by the large $Q$ values alone.
} \label{hunter_q_hisall}\end{figure}

Figure \ref{hunter_q_hisall} shows histograms of the scale heights. These heights
are more complicated to calculate than $Q$ because they require a simultaneous
equilibrium solution to the vertical distributions of both gas and stars, with the
inclusion of forces from dark matter inside the disk \citep{narayan}. A detailed
description of how this is done is in \cite{elmegreen11}. An assumed dimensionless
parameter is the ratio $\pi D$ of the column densities of dark matter to stars in
the center of the galaxy. Other dimensionless parameters are the slope of the
rotation curve, called $\alpha$ in that paper, the maximum value of the rotation
speed (at large distance) divided by the local stellar velocity dispersion, ${\hat
v}_{\rm max}$, the conventional $Q$ values for stars and gas, the ratio of the
velocity dispersions for stars and gas, and the ratio of column densities for stars
and gas. The equilibrium solution gives the central densities for stars and gas, and
then the scale heights are taken to be half of the ratios of the column densities to
the central densities. For this dimensionless calculation, the resulting scale
heights are normalized to the dimension of length, $\sigma_{\rm star}/\kappa$, which
is the half-width of a stellar epicycle in the radial direction. We find that the
scale heights do not depend noticeably on $D$, $\alpha$ and ${\hat v}_{\rm max}$,
after trying a range of values, and so we use $D=1$, $\alpha=0.4$, and ${\hat
v}_{\rm max}=8$ for what follows. For example, the value ${\hat v}_{\rm max}\sim5$
for the center of DDO 168 found by \cite{johnson15} is consistent with our
assumption. The other parameters needed to evaluate the scale heights, as listed
above, are taken from the observed radial profiles (Sect. 2).

The histograms in Figure \ref{hunter_q_hisall} show the distribution of gaseous and
stellar scale heights measured in kpc, the ratios of these heights to the local
radii, and the ratios of these heights to the local dimensional length, $\sigma_{\rm
star}/\kappa$. Evaluations of these quantities are made at each measured radius for
each galaxy. Evidently, the disks are thick in absolute terms ($H_{\rm star}\sim
H_{\rm gas}\sim0.5$ kpc), thick when compared to their local radii ($H/R\sim0.6$)
and also thick compared to their local epicyclic radii ($H\kappa/\sigma_{\rm
star}\sim3-5$).  This large relative thickness explains the finding by
\cite{dalcanton04} that dwarf galaxies do not have thin dust layers.

We conclude from Figures \ref{hunter_q_hqandscsg} and \ref{hunter_q_hisall} that the
dIrr galaxies in our sample are stable by the conventional 2D stability parameter,
$Q$, and that they are relatively thick, making them even more stable than what the
large $Q$ values imply. Still, these outer disks form stars with a nearly continuous
exponential FUV profile that approximately matches the V-band profile
\citep{hunter10}. Thus they cannot be so stable. The problem could be the
conventional 2D analysis, so we consider first in what follows a more complete 2D
model with a realistic equation of state and thickness effects included. We find
that this fails to explain the star formation also, so we follow that in Section
\ref{threed} with a consideration of the 3D problem.

\section{Stability analyses for thick two-fluid disks with turbulent energy
dissipation} \label{two-fluid}

To evaluate the possibility of gravitational instabilities in our sample galaxies,
we need to consider the calculated disk thicknesses in addition to the other
measured parameters seen in projection. The most complete model for this is the
two-fluid instability model in \cite{elmegreen11}, which also considers the
realistic case where the gas is always dissipative with a rate proportional to the
crossing time over a perturbation length.  This is expected for a supersonic
turbulent fluid, and is unlike the case for a thermal or adiabatic fluid which can
have an equilibrium unperturbed state. For a turbulent fluid, the gas always
dissipates energy unless this energy is re-supplied by either the instabilities
themselves, or by the stars that they generate. Prior to the instabilities or star
formation, however, the gas is always dissipative.

\begin{figure}\epsscale{.9}
\plotone{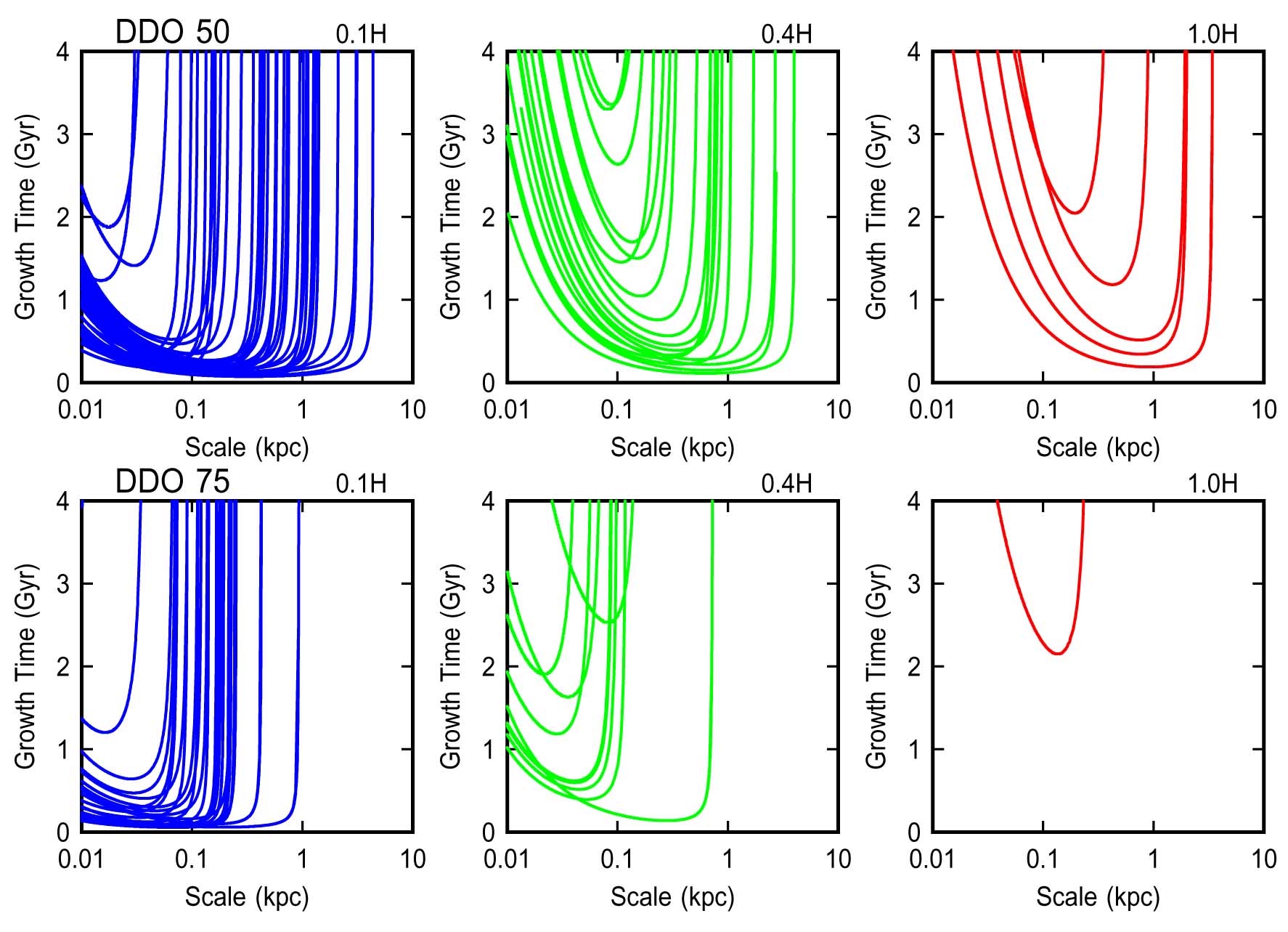}
\caption{Dispersion relations for the 2D, two-fluid gravitational instability,
plotted as unstable growth time versus pertubation
scale, with one curve for each radial interval in galaxies DDO 50 (top) and DDO 75 (bottom).
Thickness effects are included in a self-consistent way for the right-hand panels,
but for the left-hand and center panels, thicknesses equal to 0.1 and 0.4 times the true
thicknesses are assumed just to show the functional forms of these relationships. When the
thickness is forced to be smaller than the true thickness, the gravitational
instability is stronger and the growth time is smaller. These left and center
panels are unrealistic, however. At the full galaxy thickness, the growth time
of the 2D two-fluid instability is very long, indicating relatively weak
self-gravity compared to other forces.} \label{hunter_q_sigmaS_DDO50}\end{figure}

\begin{figure}\epsscale{.9}
\plotone{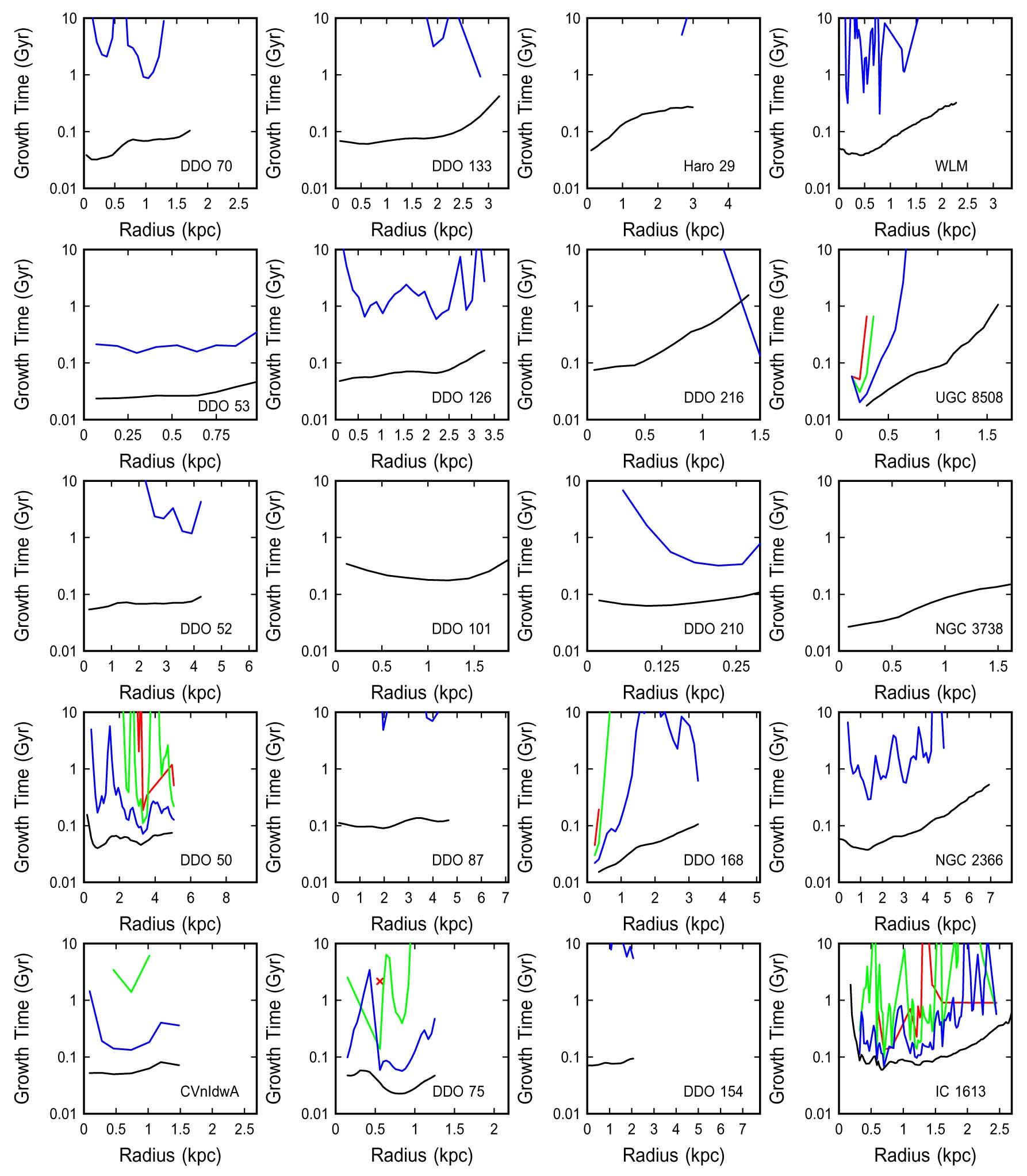}
\caption{The fastest growth times regardless of perturbation scale
for the 2D two-fluid instability are shown versus
radius for all of the galaxies in our survey. The black curves assume zero thickness
to maximize the self-gravity of the disk. Even then, the growth times are long
in some cases, equal to or exceeding 100 Myr. The blue curves assume disk thicknesses
equal to 0.1 times the true thickness, while green and red curves assume 0.4 and 1 times
the true thickness. Very few galaxies have unstable growth times less than 10 Gyr
at the full thickness, so there are almost no red curves here. }
\label{hunter_q_pkall_mix}\end{figure}

We follow the analysis in \cite{elmegreen11} and calculate the growth rate as a
function of wavenumber for each radius in each galaxy, assuming perturbations in a
dissipative star+gas mixture, using self-consistent scale heights evaluated as in
Section \ref{scaleheights}. We find that when we use the full scale heights, many
disks are so stable for all wavenumbers that the growth time exceeds a Hubble time.
To make this point, we show in Figures \ref{hunter_q_sigmaS_DDO50} and
\ref{hunter_q_pkall_mix} a selection of results for the growth time versus the
perturbation scale length (Fig. \ref{hunter_q_sigmaS_DDO50}), and the minimum growth
time considering all perturbation lengths, versus the radius (Fig.
\ref{hunter_q_pkall_mix}). The curves are color coded so that blue curves assume
much thinner gaseous and stellar scale heights for the evaluation of the two-fluid
growth time than the calculated (true) scale heights, by a factor of 10, the green
curves consider scale heights that are 0.4 times the calculated values, and the
black curves consider zero-thickness disks as an extreme case. Red curves assume the
correct thicknesses, as calculated by the method discussed in Section
\ref{scaleheights}.

In Figure \ref{hunter_q_sigmaS_DDO50}, each curve is a dispersion relation, i.e.,
growth time versus perturbation length, for a single radial annulus in the galaxy
indicated. All of the galaxies had dispersion relations like this, to varying
degrees, so we only show the results for DDO 50 and DDO 75.  Each dispersion
relation has a minimum growth time at some intermediate length scale, increasing
times for shorter lengths because these are smaller than the disk thickness, and
increasing times for longer lengths because of rotational effects. There is no
minimum scale for instability (i.e., no threshold Jeans length) in the
turbulent-dissipation model. The left-hand panels of Figure
\ref{hunter_q_sigmaS_DDO50} show the case where the thickness is forced to be 0.1
times the calculated thickness, the middle panels show 0.4 times the calculated
thickness, and the right-hand panels show the correct dispersion relation with the
full, self-consistent thickness. All radial annuli are evaluated, although for the
$0.4H$ and full-thickness cases, very few annuli have solutions for unstable growth.
This means that the actual disks are effectively stable against 2D perturbations.

The shortest growth times regardless of perturbation scale are shown in Figure
\ref{hunter_q_pkall_mix} as a function of radius for all of the galaxies. There are
very few red curves, which represent the case where the full thickness is assumed;
most of these cases had minimum growth times longer than a Hubble time and are not
shown. All galaxies have black curves plotted, which assume the thicknesses are
zero, but this is unrealistic. Some have blue curves, which are for 0.1 times the
full thickness and still unrealistic. Unstable timescales comparable to a Gyr for
$\sim0.5$ kpc perturbation scales are shown by the green curves, but even that is
somewhat unrealistic as it assumes only 0.4 times the proper thickness.

\begin{figure}\epsscale{.9}
\plotone{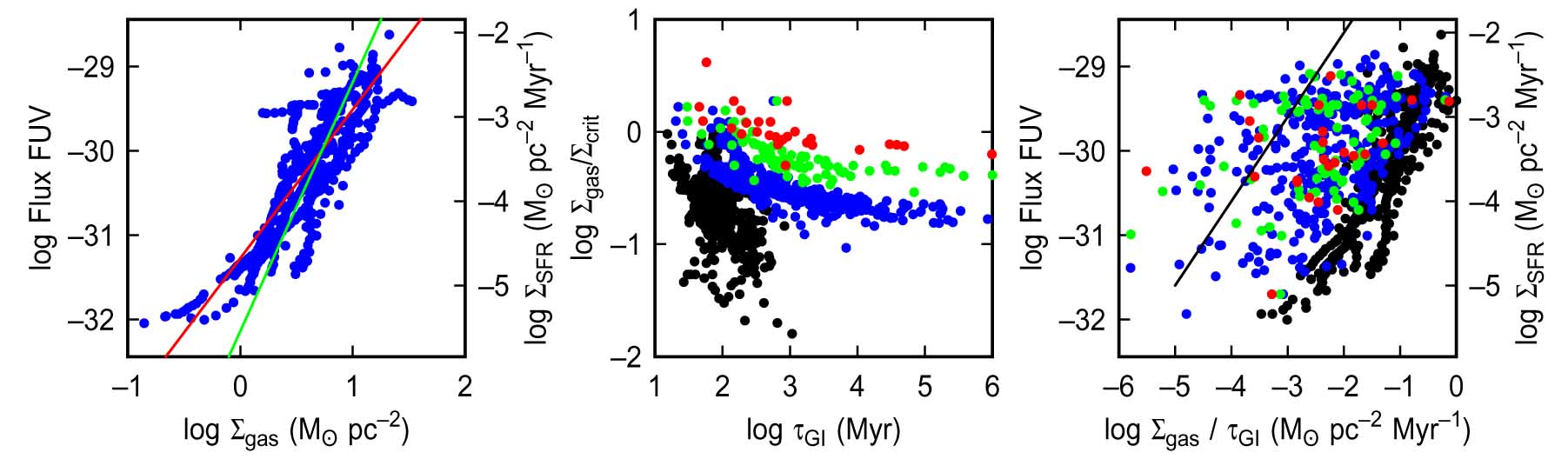}
\caption{(left) The extinction-corrected FUV flux density
in units of erg cm$^{-2}$ s$^{-1}$ Hz$^{-1}$
arcsec$^{-1}$, plotted on a $\log_{\rm 10}$ scale, is shown versus the $\log_{\rm 10}$
of the gas surface density, as determined from HI with a correction for He and
heavy elements, for all radial annuli and galaxies in our survey. The
red line is a fit to all of the points (slope $=1.76\pm0.08$)
and the green line is an
average of the fits to each galaxy (slope $=2.95\pm2.09$).
The areal SFR is on the right-hand axis. (center)
The ratio of the gas surface density to the critical value from
\cite{kennicutt89} is plotted versus the fastest growth time of the
2D two-fluid instability for the four cases of disk thickness indicated by color
(cf. Fig. 3 and 4). The red points are the most realistic
because they use the full thickness of the galaxy to determine the
self-gravitational forces.
Sub-threshold surface densities have long growth times,
too long to control star formation (i.e., they are a Gyr or
more when the full disk thickness is considered. (right)
The FUV flux and areal SFR increases with the ratio of the
gas surface density to the 2D gravitational instability time, but there is
too much scatter for all but the artificial zero-thickness case (black points)
to conclude that the 2D instability controls star formation. The black line
has a slope of 1 with arbitrary height.
} \label{hunter_q_SF_tau}\end{figure}

Figure \ref{hunter_q_SF_tau} considers the two-fluid, thick-disk result again, now
combined with measures of the SFR, taken to be proportional to the FUV flux density
as described above. First, for reference in the left panel, the FUV flux density is
plotted versus the gas surface density, showing the usual result for outer disks and
dIrr galaxies \citep{bigiel08} that star formation is proportional to the square or
higher power of the gas column density. The SFR per unit area scales directly with
the FUV flux as shown by the right-hand axis. Each point in the plot is from a
different radial annulus in one of the galaxies. The red line is a fit to all of the
points and it has a slope of $1.76\pm0.08$, where the error is the 90\% confidence
level in a Student t-test. The green line is an average of the fits to each
different galaxy; it has a slope of $2.95\pm2.09$, where the error is the standard
deviation of the individual slopes.  All of the linear fits in this paper are least
squares for ordinate versus abscissa.

The middle panel of Figure \ref{hunter_q_SF_tau} shows the ratio of the local gas
surface density to the local critical surface density for gravitational
instabilities in the usual 2D thin-disk analysis. This ratio is less than 1, as also
shown in Figure \ref{hunter_q_hqandscsg}, but now a dependence on the minimum growth
time of the local gravitational instability, $\tau_{\rm GI}$, is indicated. This is
the same two-fluid, thick-disk instability whose growth time was plotted in Figures
\ref{hunter_q_sigmaS_DDO50} and \ref{hunter_q_pkall_mix}. As the growth time
increases, the ratio $\Sigma_{\rm gas}/\Sigma_{\rm crit}$ decreases, as expected
since long growth times correspond to stable conditions. The different colors are
again for the different assumptions about thickness used for the instability
calculation (black: zero thickness, blue: 0.1H, green: 0.4H, red: full thickness).
Again, there are very few red dots and those that are plotted occur only for the
largest $\Sigma_{\rm gas}/\Sigma_{\rm crit}$ (i.e., the realistically thick disks
are stable except for the most unstable cases according to the conventional
stability criterion).

\cite{westfall14} also examined the dependence of the SFR on the two-fluid $Q$
value, finding an inverse correlation analogous to that shown here: high effective
$Q$ corresponds to longer two-fluid growth times in our figure, and thus to slower
star formation. They also find that the disks in their sample of late-type galaxies
are mostly stable because of thickness effects, like we find here.

The right-hand panel in Figure \ref{hunter_q_SF_tau} shows the FUV flux density and
$\Sigma_{\rm SFR}$ again versus a first-guess at what the SFR might be from a theory
of two-fluid, thick-disk instabilities, namely, $\Sigma_{\rm gas}/\tau_{\rm GI}$ for
unstable growth time $\tau_{\rm GI}$. The colors are as before with red showing the
realistic case of disks with their full thickness. There is no good correlation
between $\Sigma_{\rm SFR}$ and the simple 2D theory unless the disks have zero
thickness (black dots) and then the SFR is about 1\% of the theoretical rate. This
1\% is interesting because it corresponds to an efficiency of 1\% for the conversion
of gas into stars at the local dynamical rate \citep[e.g.,][]{krumholz07}. However,
the assumption of zero thickness is unreasonable. We shall recover this $\sim1$\%
efficiency in a more realistic, 3D model in the next section.

We conclude from this two-fluid, thick-disk stability analysis that our dIrr
galaxies are essentially stable against 2D self-gravitational processes. These
processes are usually considered relevant for star formation, as, for example, with
the common use of a critical column density or Toomre Q threshold, but our galaxies
have what appears to be normal star formation in these regions and yet essentially
no 2D instabilities. What this lack of 2D instability really seems to mean is that
there are no spiral waves, which are 2D processes, and that star formation is
fundamentally a 3D process. We know there are no spiral waves in these galaxies, as
they are Irregulars. The 2D gravitational instability as a spiral instability, even
considering thick disks, should be considered separately from 3D gravitational
processes that may lead to star formation.

Extrapolating a bit further, we infer that there could be an unobserved cold and
thin-disk component in these disks where star formation is actually occurring (Sect.
\ref{secondsol}). In this component, gravitational instabilities and other processes
involving self-gravity would be much stronger than in the average thick gas disk
considered above. Formation of this cold component is a 3D process since it involves
cooling, collecting and settling of thick disk gas into a thin cloudy layer. The
dynamics of star formation is probably controlled by gravitational and other forces
acting in and on this cold thin component. The Mach number could be low there, even
sub-sonic, as in the outer regions of the galaxies modeled by \cite{kraljic14}. An
example of such a cold component would be a relatively thin distribution of cool
diffuse clouds, which, because of mutual and self-shielding from background stellar
light, convert into molecular form. Mutual gravity and turbulent motions that bring
these diffuse H$_2$ clouds together could then be the triggering process for star
formation on timescales of tens of millions of years. For such a process, the
unstable growth times shown by the black curves and dots in Figures
\ref{hunter_q_pkall_mix} and \ref{hunter_q_SF_tau}, which assume zero thickness,
would be more relevant than the growth times shown by the red curves and dots, which
assume the full thickness of the HI layer given by the observed gaseous velocity
dispersions and mass column densities.

\section{A three dimensional star formation recipe}
\label{threed}

\begin{figure}\epsscale{.8}
\plotone{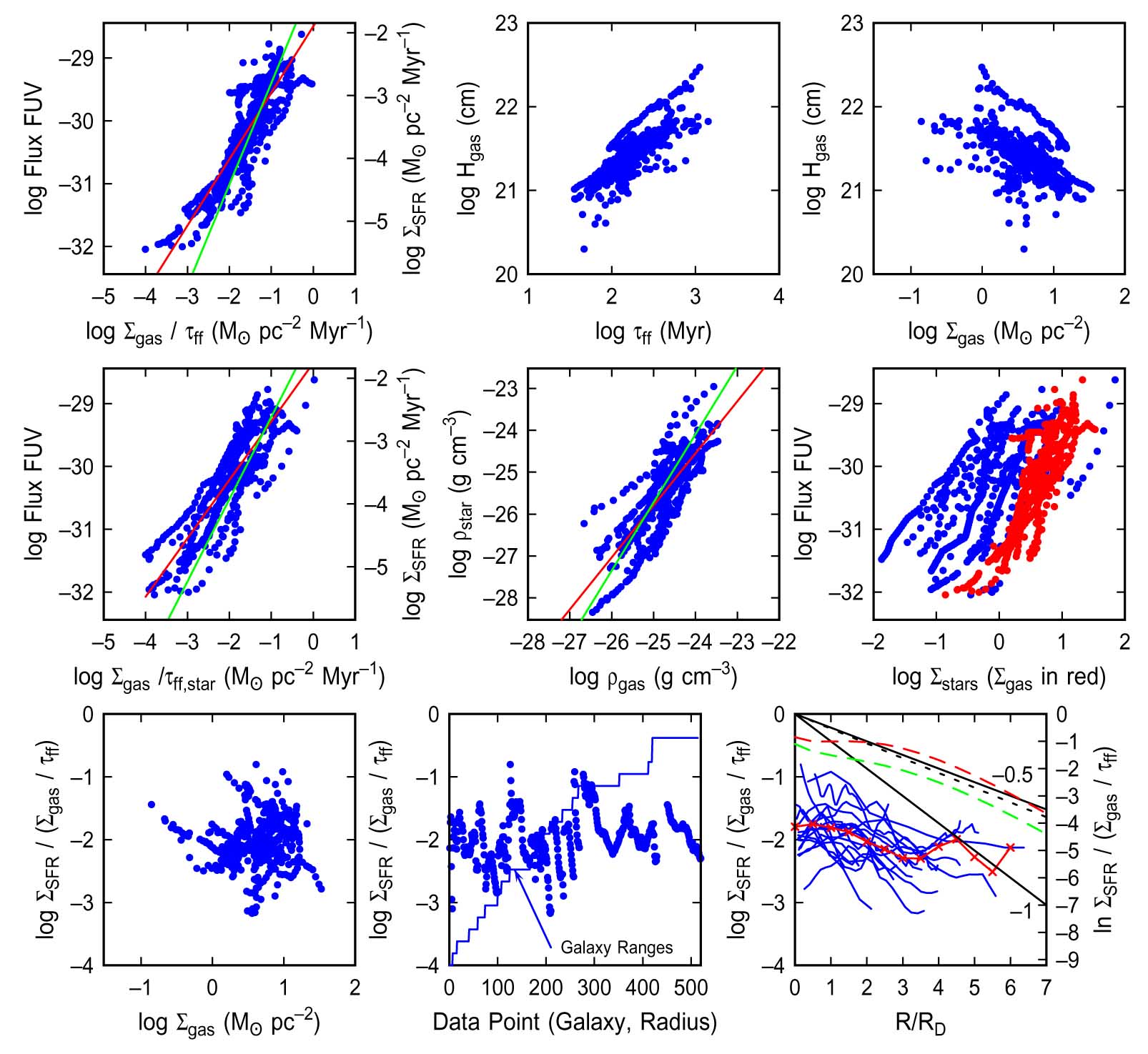}
\caption{Correlations suggesting that the areal
SFR equals an efficiency $\epsilon_{\rm ff}$
of $\sim1.0$\% (lower left) of the
HI gas surface density per unit free-fall time
at the midplane HI density, with a radial dependence of
$\epsilon_{\rm ff}$ (lower right) that parallels the dependence of the
molecular fraction in a diffuse interstellar medium (green dashed curve). Top and middle left:
SFR versus model rates using the gas and stellar midplane densities, respectively.
The correlations are similar because the stellar density is proportional
to the gas density (central panel, middle row), but it is tighter using the gas rate (top left
panel and right-hand panel in the middle row). The quadratic relation between the SFR
and the gas surface density (Fig. 5) corresponds to the upper left panel
because of the dependence of the gas scale height on the free fall time
and surface density (upper center and right panels).  In the central panel of the bottom
row, $\epsilon_{\rm ff}$ is plotted versus radius for all of the galaxies from left
to right, as indicated by each horizontal segment in the rising curve. A radial
decrease is evident for each one, as also shown in the lower right panel
where each galaxy is a different curve with an average slope given
by the dotted black line.  The other two lines in the lower right have
fiducial slopes of $-0.5$ and $-1$ for comparison. The red dashed curve is the
molecular fraction when the radial profile of FUV light is used to determine
the dissociation rate. This red-dash curve is not a good fit to the data
in these cases where it is
assumed that the H$_2$ and HI thicknesses increase with radius in the same way.
The FUV model is a better fit if the H$_2$ thickness is constant, as
discussed in Sect. \ref{secondsol}.
} \label{hunter_q_SF_radial}\end{figure}

The measured surface densities and derived disk thicknesses allow us to calculate
the 3D densities for both gas and stars, and therefore the Jeans time, $(4\pi G
\rho)^{-0.5}$, or the idealized free-fall time for a uniform sphere, $\tau_{\rm
ff}=(32 G \rho/3\pi)^{-0.5}$. Figure \ref{hunter_q_SF_radial} shows various
comparisons between the SFR and models using $\tau_{\rm ff}$. These are now 3D
models because they involve space density for the dynamical rates, not solutions to
the 2D growth rates.

The upper left panel of Figure \ref{hunter_q_SF_radial} shows the FUV intensity and
$\Sigma_{\rm SFR}$ versus a 3D first-guess model for the SFR, $\Sigma_{\rm
gas}/\tau_{\rm ff}$, using the observed HI properties for the gas. There is a good
correlation with slopes of $1.06\pm0.04$ in the red line, which is for all of the
points together, and $1.61\pm0.58$ in the green line, which is the average of all of
the slopes for the individual galaxies without any weighting. The errors are as
above, from a Student t-test and from the rms scatter in the individual slopes,
respectively. The closeness of these slopes to unity implies the surface SFR is
tracking the 3D dynamical rate.

The conversion between an approximately squared dependence of $\Sigma_{\rm SFR}$ on
$\Sigma_{\rm gas}$ and an approximately linear dependence of $\Sigma_{\rm SFR}$ on
$\Sigma_{\rm gas}/\tau_{\rm ff}$ implies that the disk thickness is changing
systematically with both $\tau_{\rm ff}$ and $\Sigma_{\rm gas}$. These dependencies
are shown in the middle and right-hand panels on the top of Figure
\ref{hunter_q_SF_radial}, where $H_{\rm gas}\propto \tau_{\rm ff}$ and $H_{\rm
gas}\propto \Sigma_{\rm gas}^{-1}$. Note that for equilibrium in an isothermal
one-component fluid, $H=\sigma^2/(\pi G \Sigma)$ and $\rho=\Sigma/(2H)$. If
$H\propto \Sigma^{-1}$ as in the upper right-hand panel, then $\rho\propto\Sigma^2$
and $\tau_{\rm ff}\propto\Sigma^{-1}$. This converts $\Sigma^2$ as in Figure
\ref{hunter_q_SF_tau} left, to $\Sigma/\tau_{\rm ff}$ as in Figure
\ref{hunter_q_SF_radial} top-left. In fact Figure \ref{hunter_q_SF_radial} uses fits
to the HI thickness involving gravity from gas, stars and dark matter, but the
scaling between quantities ends up about the same as in this idealized isothermal
model. The physical point is that the empirical star formation law appears to change
from approximately linear with gas surface density in the inner parts of spiral
galaxies to approximately quadratic in the outer parts of spiral galaxies and in
dIrrs because of a general thickening of the disk at low surface densities. This
point was also made by \cite{barnes12} using observations of spiral galaxies.

The left panel in the middle row of Figure \ref{hunter_q_SF_radial} shows the FUV
intensity and $\Sigma_{\rm SFR}$ versus a similar model SFR where now the 3D average
local HI gas density used for the free fall time has been replaced by the 3D average
local stellar density. The motivation for this replacement comes from observations
of the good correlation between the SFR and the stellar surface density
\citep{shi11} or stellar volume density \citep{ostriker10,kim13} (see also Fig.
\ref{hunter_q_shi} below).  The fitted lines indicated have slopes of $0.93\pm0.04$
(all points, red line) and $1.31\pm0.56$ (average of individual galaxies, green
line), with the error limits defined as before. The correlation is about as good as
when the free fall time involves the HI gas density, although there is a little more
horizontal scatter in the stellar density case. This similarity for the two cases is
because the stellar and HI gaseous densities scale with each other (middle panel,
middle row). For this density-density correlation, the indicated fits have slopes of
$1.24\pm0.07$ (all points, red line) and $1.63\pm0.84$ (average of individual
galaxies, green line). The reason the stellar density scales steeper than linearly
with the gas density is that the gas-to-star ratio tends to increase in the outer
parts of the disks. Indeed, at the lowest values in the outer disks, the HI gas
densities are about 10 times the stellar densities; at the highest values in the
inner disks, the HI gas and star densities are about the same.

The right-hand panel in the middle row of Figure \ref{hunter_q_SF_radial} plots the
FUV intensity versus the gas (red) and stellar (blue) surface densities. The gas
values are clearly shifted towards higher values. The dispersion in the correlation
is higher for the stars than the gas, suggesting that the physical processes of star
formation are determined mostly by the gas, with stars playing a secondary role,
such as determining the scale height, radiation field, etc.

The bottom-left panel shows the ratio of the observed SFR, calculated as before from
the azimuthally-averaged FUV intensity, to the 3D first-guess SFR, calculated as
before from the ratio of the azimuthally-averaged HI gas surface density to the 3D
free fall time. This ratio has become known as the efficiency of star formation per
unit free-fall time, $\epsilon_{\rm ff}$, considering a star formation law where the
SFR per unit area equals $\epsilon_{\rm ff}\Sigma_{\rm gas}/\tau_{\rm ff}$
\citep{elmegreen02,krum09,krumholz12}. All of the points cluster around a nearly
constant efficiency that is within a factor of three of 1\%. The actual average in
the figure is $\log_{10}\epsilon_{\rm ff}=-1.99$, which corresponds to
$\epsilon_{\rm ff}=1.0$\%.

The efficiency is shown again in the middle panel of the bottom row of Figure
\ref{hunter_q_SF_radial}, with one point for each radial annulus in each galaxy, but
now spread out along the abscissa so that radial trends can be seen. The plot moves
from left to right with increasing radius and from one galaxy to the next.  There is
an increasing segmented line showing the jumps from one galaxy to another; i.e., the
segmented line is horizontal inside each galaxy and then increments upward by one
unit for the next galaxy. The order of the galaxies is the same as in Table 1. The
scatter in efficiency in the bottom-middle panel is the same as in the bottom-left
panel, as the ordinate values are exactly the same, but now a trend appears such
that within each galaxy, the points show a systematic decrease with increasing
radius.

These radial decreases in efficiency are shown again in the bottom-right panel where
each galaxy is now plotted as a separate curve, normalized in radius to the V-band
scale length, $R_{\rm D}$. There is clearly a systematic dependence of
$\epsilon_{\rm ff}$ on $R/R_{\rm D}$; the average trend is given by the red curve.
The left-hand axis for this plot has the log of $\epsilon_{\rm ff}$ with a base-10
logarithm, while the right-hand axis has $\ln \epsilon_{\rm ff}$ with a natural log.
The two black straight lines in the bottom-right panel are drawn to guide the eye;
they have slopes of $-1$ and $-0.5$ using the natural log axis, as labeled. The
average fitted slope of $-0.54$ is given by the black dotted curve. The trend in
$\ln \epsilon_{\rm ff}$ versus $R/R_{\rm D}$ has a slope of about $-0.5$, which
means that $\epsilon_{\rm ff}$ decreases exponentially with radius in proportion to
the square root of the V-band surface brightness, which would have a slope of $-1$
on such a plot (e.g., see Figure \ref{hunter_q_sr_radials} below). In other words,
$\epsilon_{\rm ff}$ is exponential with a scale length that is about twice the scale
length of the optical disk. An exponential decrease in the efficiency with radius
was also shown by \cite{yim14} for four spiral galaxies. In the next section, we
discuss a possible origin for the radial dependence of the efficiency of star
formation.

The dependence of the SFR on the surface density of stars deserves another look.
\cite{shi11} find a good correlation with $\epsilon_{\rm ff}\propto\Sigma_{\rm
star}^{0.5}$.  We show the analogous dependency for our galaxies in Figure
\ref{hunter_q_shi}. This differs from the left panel in the middle row of Figure
\ref{hunter_q_SF_radial}, because that uses $1/\tau_{\rm ff}\propto\rho_{\rm
star}^{0.5}$  on the abscissa instead of $\Sigma_{\rm star}$. The correlation in
Figure \ref{hunter_q_shi} is steeper than the Shi et al. correlation, with an
average slope of 0.76, as shown by the dashed red line. However, if we consider only
the low surface brightness and late-type galaxies in \cite{shi11}, which are most
similar to our galaxies, then their correlation with $\Sigma_{\rm star}$ is also
steeper than their derived average slope of 0.5. Consider the LSB and late-type
points in their Figures 1 and 3, and especially their Figure 5, which is the
combined result for 12 of their spiral galaxies. Their plotted points for those
galaxies, ie., for the lowest SFR galaxies in their Figure 5, have a steeper slope
than all of the points combined, and steeper than their least-squares fit to all of
the points. The slope for the low-SFR points in their Figure 5 is closer to 1 than
0.5, and in approximate agreement with the slope we find here in Figure
\ref{hunter_q_shi}. Thus our data agree approximately with the THINGS data used in
\cite{shi11}.

\section{Radial variation of the efficiency as an indicator of varying molecular fractions}
\subsection{Radial variations}
\label{radial}

The 3D star formation model used in the previous section,
\begin{equation}
\Sigma_{\rm SFR}(R) = \epsilon_{\rm ff}(R) \Sigma_{\rm gas}(R) /\tau_{\rm ff}(R),
\label{sfreq}
\end{equation}
contains the quantity $\Sigma_{\rm gas}$ that has been set equal to the HI column
density. Considering the measured value of $\epsilon_{\rm ff}\sim1$\%, the
implication of this equation is that approximately 1\% of the HI gas turns into
stars per unit free fall time as measured at the average midplane density of HI. Of
course, stars are not forming at the average midplane HI density; they form in
denser clouds. Nor is the free fall time in the star-forming gas as long as the free
fall time at the average density; stars form in dense molecular gas on much shorter
local timescales (discussed below in Sect. \ref{secondsol}).

To get some sense of the values and radial trends for the gaseous and stellar
surface densities, 3D space densities, and scale heights, Figure
\ref{hunter_q_sr_radials} plots these quantities using a separate curve for each
galaxy.  The plots use a natural logarithm on the ordinate and a linear normalized
radial scale on the abscissa, so that the slopes of the curves can be readily
identified in relation to the exponential light profiles, which all have slopes of
$-1$, by definition. Each panel also shows fiducial slopes of $-1$ and $-0.5$ as
straight lines, and in the cases of 3D stellar density and FUV flux, $-1.5$, as
labeled. The average trends are summarized in Table 2, which gives the natural log
of the central value (``V'') for comparison to the figure, the physical values at
the center and at $4R/R_{\rm D}$, and the exponential factor, which is the slope in
the figure. Standard deviations come from the galaxy-to-galaxy variation. Table 2
also gives the average radial variation of gas velocity dispersion from the HI
observations and the derived quantity $\tau_{\rm ff}$.

There are clear trends shared by these galaxies. The stellar surface density profile
(top middle) has a slope of about $-1$, showing essentially the exponential disk as
viewed in the V-band (although the surface density was determined from SED fitting).
The stellar and gaseous scale heights (top and middle right-hand panels) are similar
to each other and both increase with radius about as $\exp(0.5 R/R_{\rm D})$. The
stellar volume density therefore decreases with radius as $\exp(-1.5R/R_{\rm D})$,
from $\Sigma_{\rm star}/H_{\rm star}$, as shown in the top left panel.  This rapid
decrease of stellar density corresponds to a similar decrease in volume emissivity
for V-band stellar light, $j_{\rm V}$, and it occurs throughout the entire radial
range in most of the disks; volume emissivity does not follow the projected radial
profile but is steeper because of the steady increase with radius in the disk
thickness.

The surface density of HI gas follows a shallower trend than the surface density of
stars, being more like $\exp(-0.5R/R_{\rm D})$. Thus the 3D HI density varies as
$\Sigma_{\rm gas}/H_{\rm gas}\propto \exp(-R/R_{\rm D})$, as shown in the middle
left panel. The absorption coefficient from dust, $\kappa_{\rm a}$, is proportional
to the gas density for constant metallicity \citep[as observed in dIrrs;][]
{pagel80,roy96,kobul96,kobul97}, so it should also follow the $\exp(-R/R_{\rm D})$
trend. As a result, the local radiation field in V-band, $<\Phi_{\rm V}>$ does not
mimic the projected light distribution, which would be $\exp(-R/R_{\rm D})$, but
rather $<\Phi_{\rm V}>\sim j_{\rm V}/\kappa_{\rm a}\sim\exp(-0.5R/R_{\rm D})$, which
has twice the scale length.  Local radiation fields are higher in the outer parts of
these galaxies than the projected light profiles suggest.  Similarly, the local gas
density drops faster with radius than the observed column density.

On the bottom left of Figure \ref{hunter_q_sr_radials} is the FUV flux density and
$\Sigma_{\rm SFR}$ versus radius in a plot of the natural-log versus $R/R_{\rm D}$.
This panel shows a variation of the areal SFR that is proportional to the power
$1.5$ of the V-band surface brightness, i.e., the slope is $-1.5$ on that plot. Thus
the volume emissivity of the SFR scales as $\exp(-2R/R_{\rm D})$, considering the
$H_{\rm star}$ variation. In the bottom central panel is the ratio $n_{\rm
star}/n_{\rm gas}$, which is proportional to the ratio of the V-band volume
emissivity, $j_{\rm V}$, to the absorption coefficient from dust, $\kappa_{\rm a}$;
it scales with the square root of the V-band surface brightness.

This difference between local radiation field and surface brightness calls into
question an assumption of the star formation models in \cite{ostriker10},
\cite{kim13}, and \cite{krumholz13}, which was that the local radiation field
follows the SFR per unit area. As mentioned above, the radiation field in V-band is
approximately $<\Phi_{\rm V}>\sim j_{\rm V}/\kappa_{\rm a}$ where $j_{\rm
V}\propto\Sigma_{\rm star}/H_{\rm star}$ and $\kappa_{\rm a}\propto \rho_{\rm
gas}\propto\Sigma_{\rm gas}/H_{\rm gas}$ for small metallicity gradient. Thus
$<\Phi_{\rm V}>\propto \Sigma_{\rm star}H_{\rm gas}/\left(\Sigma_{\rm gas}H_{\rm
star}\right)$. But $H_{\rm gas}\propto H_{\rm star}$ from Figure
\ref{hunter_q_sr_radials}, thus $<\Phi_{\rm V}>\propto\Sigma_{\rm star}/\Sigma_{\rm
gas}$ which is shallower than $\Sigma_{\rm star}$ alone.  Similarly, in FUV,
$<\Phi_{\rm FUV}>\propto j_{\rm FUV}/\kappa_{\rm a}\propto\exp(-R/R_{\rm D})$ which
is shallower than $\Sigma_{\rm SFR}\propto\exp(-1.5R/R_{\rm D})$. Also in a study of
17 spirals and 5 dIrrs in the THINGS survey \citep{bigiel10}, $\Sigma_{\rm gas}$ and
the intensity of FUV both vary approximately exponentially in the outer parts, with
a generally shallower slope for $\Sigma_{\rm gas}$ than FUV; thus the local
radiation field should be shallower than the FUV profile for the THINGS galaxies
too.

The bottom right panel of Figure \ref{hunter_q_sr_radials} shows the natural
logarithm of the ratio $\Sigma_{\rm gas}/\tau_{\rm ff}$ versus normalized radius.
There is a direct scaling with the V-band (i.e., the slope is about $-1$). This
direct scaling also follows from the middle panels Figure \ref{hunter_q_sr_radials}
considering that $\tau_{\rm ff}\propto\rho_{\rm gas}^{-1/2}$, because in the
middle-center panel, $\Sigma_{\rm gas}\propto \exp(-0.5R/R_{\rm D})$ and in the
left-center panel, $\rho_{\rm gas}\propto\exp(-R/R_{\rm D})$.  The ratio of the
plotted quantities in the bottom left and right-hand panels is the efficiency,
plotted in the lower right of Figure \ref{hunter_q_SF_radial}. This ratio has the
residual dependence on $\exp(-0.5R/R_{\rm D})$ as shown in Figure
\ref{hunter_q_SF_radial} because the SFR profile is steeper than the purely
dynamical profile given by $\Sigma_{\rm gas}/\tau_{\rm ff}$ for HI. The residual
variation could be from the molecular fraction, as discussed in the next section.

We note that the radial variation of the disk density from stars and dark matter,
usually dominated by stars, is not considered in \cite{krumholz13}, which is a study
of star formation at low HI surface density, as is the present paper. This quantity,
called $\rho_{\rm sd}$ in that paper, is taken to be constant for all gas surface
densities, $\Sigma$, in their theoretical Kennicutt-Schmidt plots of $\Sigma_{\rm
SFR}$ versus $\Sigma_{\rm gas}$, when in fact all three quantities, $\Sigma_{\rm
SFR}$, $\Sigma_{\rm gas}$ and $\rho_{\rm star}$ vary with radius for the galaxies
studied here.  That is, $\rho_{\rm star}$ varies as approximately the 3rd power of
$\Sigma_{\rm gas}$ in the radial direction.

An interesting feature of the radial dependencies for $\Sigma_{\rm gas}(HI)$ and
$n(HI)$ is that $n(HI)$ is proportional to the total interstellar pressure, $P$.
This is because $P\sim\pi G\Sigma_{\rm gas}^2/2$ for our gas-dominated disks and
$\Sigma_{\rm gas}\propto\exp(-0.5R/R_{\rm D})$ from Figure
\ref{hunter_q_sr_radials}. Thus $P\propto\exp(-R/R_{\rm D})$, and this is also
proportional to $n(HI)$, from Figure \ref{hunter_q_sr_radials}.

\subsection{Molecular diffuse clouds}
\label{mdc}

The average midplane HI gas densities in our dIrr disks are rather low, with values
of $\exp(-1.5)=0.2$ cm$^{-3}$ starting in the center and then decreasing outward
(middle row, left panel in Fig. \ref{hunter_q_sr_radials}).  This implies that the
gas is generally tenuous, much more so than in the local solar neighborhood. Still,
the gas is probably cloudy with the observed HI only an average over clouds inside
the HI resolution limit and spread out around a radial annulus.  The denser parts of
these clouds and those shielded from radiation could be partially molecular,
although not necessarily self-gravitating. We refer to non-self-gravitating clouds
as diffuse.

``Standard'' diffuse clouds in the solar neighborhood are mostly H$_2$
\citep{hollenbach71,spitzer75,jura75}; CO is not observed from them in emission
although some CO is observed in absorption \citep{federman}.  To some extent, clouds
in the solar neighborhood shield each other from H$_2$ dissociating radiation; to
them, a large fraction of the sky is nearly black in the Lyman-Werner bands unless
an early type star is closer than the standard cloud mean free path, which is about
100 pc. The top panel of Figure \ref{figure9_china2012} shows a typical spectrum of
a nearby hot star in an H$_2$ absorption line, from \cite{spitzer75}. The line
center flux is effectively zero, so no photo-dissociating radiation from this
transition gets through the intervening clouds to us.

Evidence for diffuse H$_2$ was shown in a study of dust and HI emission from the
Perseus region (Lee et al. 2012, see also Barriault et al. 2010). The middle panel
of Figure \ref{figure9_china2012} illustrates this region of the sky and the bottom
panel shows the probability distribution function (pdf) of K-band extinction, both
from \cite{lombardi10}. Superposed on the extinction distribution is the threshold
for H$_2$ formation determined by the Copernicus satellite in the 1970's
\citep{spitzer75}, namely, $A_{\rm V}>0.3$ mag for the solar neighborhood ($A_{\rm
K}>0.033$ mag in K-band). This extinction threshold corresponds to a mass surface
density threshold $\Sigma \sim 5.9 \;M_\odot$ pc$^{-2}$, and is reproduced by the
theory of H$_2$ formation
\citep[e.g.,][]{hollenbach71,jura75,krumholz09b,sternberg14}. Also shown in the
figure is the threshold for CO formation, which is approximately $A_{\rm V}\gtrsim
1.5$ mag ($A_{\rm K}>0.165$ mag) in the solar neighborhood
\citep{federman,pineda08}. Note the gray-scale bar on the right in the middle panel.
The blue horizontal bar near the bottom of it indicates the gray scale threshold for
H$_2$ formation.  Most of the image is darker than this, suggesting that molecular
diffuse clouds nearly cover the field. Considering also the CO threshold, a small
fraction of the area is likely to contain CO, although most of it contains H$_2$.

A similar situation but more extreme compared to that in the solar neighborhood
arises in dwarf Irregular galaxies where low metallicities make CO hard to form and
detect but significant amounts of H$_2$ are inferred anyway from a high abundance of
dust without the usual accompaniment of HI (Leroy et al. 2008; Hunt et al. 2010;
Bolatto et al. 2011; Elmegreen et al. 2013; Shi et al. 2014; see review in Bolatto
et al. 2013).

The formation of H$_2$ in giant molecular clouds (GMCs) can be more easily
understood if GMCs are collections of diffuse H$_2$ clouds. If the H$_2$ forms
before the GMCs, then there is no problem with slow molecule formation like that
discussed by \cite{maclow12}. The gas could spend a relatively long time in the
diffuse phase forming $H_2$ if the conditions are right, and then convert back to HI
after the GMC phase, when star formation breaks the clouds apart. Molecule formation
should not only be a {\it self}-shielding process for individual clouds, but also a
{\it mutual} shielding process for collections of clouds shielding each other
\citep{elmegreen93}. Other discussions of diffuse molecular clouds are in
\cite{elmegreen13} and \cite{shetty13,shetty14}.

For reference, the local threshold column density for star formation discussed by
\cite{lada10} and \cite{evans14} is also shown in Figure \ref{figure9_china2012}. It
is $A_{\rm K}>0.8$ mag, or $A_{\rm V}>7.3$ mag, or $\Sigma_{\rm gas}> 145\;M_\odot$
pc$^{-2}$ including He and heavy elements at solar metallicity.

The low gas densities in our dIrr sample suggest that most of the interstellar
medium is a mixture of warm and cool HI, with a non-negligible but small fraction of
the mass in molecular form. Most of these molecules are probably diffuse, also
because of the low average density, with sparsely distributed cores in which gravity
is strong and stars form. In that case, the molecular fraction can be determined in
some average sense by an equilibrium between H$_2$ formation at the root mean
squared density, and H$_2$ destruction at the mean density in the local average
radiation field. This is a different model than determining the transition from
atomic to molecular gas at the edge of a single dense cloud, and also different from
determining the average molecular fraction from the fraction of the gas mass in the
form of isolated dense molecular clouds, which may have applications elsewhere
\citep[e.g.,][]{krumholz09b}. We return to this point in Section \ref{sharp}.

The equilibrium may be written approximately as
\begin{equation}
n(HI)n({\rm dust})R_{\rm form} = n(H_2)f_{\rm diss}\Phi\sigma_{\rm dis}
\label{formation}
\end{equation}
where $R_{\rm form}$ is the formation rate of molecules on dust
\citep[e.g.,][]{hollenbach71,jura75}, $\Phi$ is the local radiation field and
$\sigma_{\rm dis}$ is the cross section for H$_2$ dissociation in the radiation
field (which includes the fraction of absorptions leading to dissociation). Equation
\ref{formation} is analogous to equations (37)-(38) in \cite{mckee10}, who consider
an additional average over radiation frequency. If $n(H_2)<<n(HI)$ on average, then
\begin{equation}
{{n(H_2)}\over{n(HI)}} \sim {{ n(HI)D_{\rm DTG}R_{\rm form}}
\over{\Phi\sigma_{\rm dis}}} \propto
{{n(HI)(Z/Z_\odot)}\over{\Phi}}
\label{meanmole}
\end{equation}
for dust-to-gas ratio $D_{\rm DTG}$ and metallicity dependence $Z$ from the dust. We
expect that the efficiency of star formation, $\epsilon_{\rm ff}$, contains a
dependence on $n(H_2)/n(HI)$, which effectively converts the $\Sigma_{\rm gas}$ from
HI observations into a local molecular column density. This is because most star
formation occurs in molecular gas even at low average density and moderately low
metallicity \citep{wong02,krumholz11}. Since we measure the radial distribution of
$n(HI)$, we need to evaluate the radial distribution of the radiation field $\Phi$
in order to compare the expected molecular fraction from equation (\ref{meanmole})
with the observation of $\epsilon_{\rm ff}(R)$. A more complete theory would also
consider the distribution function of diffuse cloud column densities and the
discrete nature of the brightest radiation sources.

Equation (\ref{meanmole}) is similar to the result in \cite{elmegreen93} where,
following a more detailed model, the molecular fraction at low molecular fraction,
in a combined diffuse and self-gravitating medium, was written as approximately
$P^{2.2}/j$ for interstellar thermal pressure $P$. The similarity arises because
$P\propto n(HI)$ for constant diffuse cloud temperature and because volume
emissivity $j\propto \Phi\kappa_{\rm a}\propto \Phi n(HI)$ for constant metallicity.

\subsection{Radiative transfer in the diffuse interstellar medium of dIrr disks}
\label{radtrans}

We would like to evaluate the radial distribution of the radiation field in our dIrr
galaxies, given the observed and derived radial distributions of the HI and stellar
densities and the derived scale heights.  We do this by solving the integral form of
the equation of radiative transfer,
\begin{equation}
I(\theta,\phi,R_{\rm O})=\int_0^{s_{\rm max}} j(s) e^{-\tau(s)}ds
\label{radt}
\end{equation}
where optical depth is
\begin{equation}
\tau(s)=\int_0^s \kappa_{\rm a} (s^{\prime}) ds^{\prime}
\label{radtau}
\end{equation}
for variables $s$ and $s^\prime$ along a sightline in the direction $(\theta,\phi)$,
where $\theta$ is the polar angle with respect to the $z$ direction at the position
of the observer (subscript ``O'') at galactocentric radius $R_{\rm O}$, and $\phi$
is the azimuthal angle measured by the observer around the disk starting at the
direction to the galactic center. The volume emissivity and absorption coefficient
depend on position in the galaxy. For the V-band emission,
\begin{equation}
j_{\rm V}(R,z)=\exp(-1.5R/R_{\rm D}-z/H_{\rm star})\;\;,\;\;
\kappa_{\rm a}(R,z)=\kappa_0\exp(-R/R_{\rm D}-z/H_{\rm gas}).
\end{equation}
Here we have used two profiles from Figure \ref{hunter_q_sr_radials}, the HI density
profile for the absorption coefficient and the stellar density profile from the
V-band.  The HI density profile is a good representation of the absorption
$\kappa_{\rm a}$ if the dust is not confined to dense tiny clouds (much smaller than
the HI angular resolution). Such dense clouds can have a long physical mean free
path. The ratio of the physical mean free path between clouds to the photon mean
free path for dust absorption, $1/\kappa_{\rm a}$, is the intrinsic dust opacity of
each cloud, $\tau_{\rm cl}$.  If the intrinsic opacity is small, then photons travel
through many clouds before they get absorbed on dust and the average density from HI
is relevant for $\kappa_{\rm a}$.  However, the use of $\kappa_{\rm a}\propto n(HI)$
does not necessarily assume that all of the absorption is by dust.  The real
assumption is that the average radiation field in a local volume containing many
clouds varies with galactocentric radius as if the total opacity from dust and
molecules is proportional to $n(HI)$. The molecular fraction and equation
(\ref{formation}) then refer to average formation and dissociation rates among all
clouds inside this local volume. It could still be that some clouds are more
molecular than others depending on density and opacity.

We also consider a second model that uses the FUV volume emissivity profile $j_{\rm
FUV}\propto\exp(-2R/R_{\rm D}-z/H_{\rm star}$) from the FUV flux density in Figure
\ref{hunter_q_sr_radials} divided by $H_{\rm star}$. The FUV is more appropriate for
H$_2$ dissociation, but if other factors are important, such as thermal heating,
cosmic ray flux, or stellar mass, then the V-band profile might be relevant too.

We are interested in the radial trends, so the volume emissivities are normalized to
the values at the galactic center. The scale heights are also assumed to vary with
radius, as observed in Figure \ref{hunter_q_sr_radials},
\begin{equation}
H_{\rm gas}\sim400e^{0.5R/R_{\rm D}} \;{\rm pc}\;\;;\;\;
H_{\rm star}\sim400e^{0.5R/R_{\rm D}}\;{\rm pc}.
\end{equation}
The central absorption coefficient, $\kappa_{\rm a}(R=0)$, is given by the relation
\begin{equation}
2H_{\rm gas}(R=0)\kappa_{\rm a}(R=0)=\ln(10^{0.4A_{\rm V}[R=0]})
\end{equation}
where the visual extinction in the center, $A_{\rm V}(R=0)$, comes from the total
hydrogen column density there, $N$, as $A_{\rm V}=N/1.87\times10^{21}$, using
$E(B-V)=N/5.8\times10^{21}$ from Bohlin et al. (1978) with a ratio of total to
selective extinction $R=3.1$. Considering that $A_{\rm V}/N$ should scale with
metallicity $Z$,
\begin{equation}
A_{\rm V}(R=0)=\Sigma_{\rm gas}(R=0)(Z/Z_\odot)/(1.87\times10^{21}\mu).
\end{equation}
We evaluate $\kappa_{\rm a}$ at the center assuming $\ln\Sigma_{\rm gas}=2$ at the
center, from the center-middle panel of Figure \ref{hunter_q_sr_radials}, and using
a mean molecular weight of $\mu=2.2\times10^{-24}$ g. We also assume $Z/Z_\odot=1/8$
which is about the average for our dIrr galaxies. This gives a central photon mean
free path for visible light of $1/\kappa_{\rm a}=18.4$ kpc. Considering also that
$A_{\rm FUV}=8.24E(B-V)=2.6A_{\rm V}$ from \cite{Cardelli_etal89}, the central
absorption coefficient in FUV corresponds to $1/\kappa_{\rm FUV}=7.1$ kpc. The
radial scale length in V-band, $R_{\rm D}$, is taken equal to the average for our 20
galaxies, which is 740 pc, using the photometry and distances in the LITTLE THINGS
survey \citep{hunter12}.

The galactic cylindrical coordinates $(R,z)$ are related to $s$, $\theta$, and $\phi$
from the point of view of the observer by the equations
\begin{equation}
R=\left(\left[s \sin\theta\cos\phi-R_{\rm O}\right]^2
+\left[s \sin\theta\sin\phi\right]^2\right)^{0.5}\;\;;\;\;z=s\cos\theta .
\end{equation}
This is enough to give the intensity in all directions at radius $R_{\rm O}$. The
average radiation field is then the integral,
\begin{equation}
\Phi(R_{\rm O})={1\over{4\pi}}
\int_{-\pi}^{\pi} \sin\theta d\theta\int_0^{2\pi}d\phi I(\theta,\phi,R_{\rm O}).
\label{radphi}
\end{equation}
The integrals (\ref{radt}), (\ref{radtau}), and (\ref{radphi}) were evaluated
numerically with enough resolution in all coordinates $s$, $\theta$, and $\phi$,  to
obtain convergence. The integral over the line of sight, $s$, had a limit in
galactocentric radius, $R$, of 20 scale lengths and a limit in the vertical
direction, $z$, of 6 times the local scale height for stars.

The results of this integral are used for the evaluation of mean molecular fraction
in equation (\ref{meanmole}), where $\Phi$ occurs in the denominator beneath the gas
density $n(HI)$. If the residual radial trend in the star formation efficiency per
unit free fall time, $\epsilon_{\rm ff}$, is the result of variations in the mean
molecular fraction, then the curves in the lower right of Figure
\ref{hunter_q_SF_radial} should follow the trend given by equation (\ref{meanmole}).
Thus we plot $n(HI)/\Phi$ in that figure. The green dashed curve is for V-band light
with $j_{\rm V}\propto\exp(-1.5R/R_{\rm D})$ and $1/\kappa_{\rm a}(R=0)=18.4$ kpc,
and the red dashed curve is for FUV light with $j_{\rm FUV}\propto\exp(-2R/R_{\rm
D})$ and $1/\kappa_{\rm FUV}(R=0)=7.1$ kpc. The curves are scaled arbitrarily in the
vertical direction. The FUV curve is flatter than the V-band curve because the
volume emissivity is steeper in FUV than in V-band. Both curves are nearly flat in
the center because the mean free path there is larger than the radius, so an
observer would see through to the other side of the galaxy where the emissivity is
low.

The radiative transfer solution in Figure \ref{hunter_q_SF_radial} follows the trend
of the efficiency, $\epsilon_{\rm ff}$ when the V-band light is used for the
molecular fraction (green dashed curve). In this case, the SFR is proportional to
the product of three quantities: the available HI gas mass, the 3D gravitational
rate measured at the midplane density of HI, and the diffuse molecular fraction,
which is assumed proportional to the ratio of the gas density to the V-band
radiation field. This star formation law is fairly simple, but dIrr galalies may be
simple too: gas dominates stars in mass, extinction is weak, the molecular fraction
may be low, and the disk is too thick to support 2D instabilities like spiral waves.
The dependence on 3D density does not indicate a particular star formation
mechanism; many mechanisms have a characteristic timescale proportional to the
average self-gravity time, including gas collapse in expanding shells, collapse
following turbulence compression, and 3D gravitational instabilities in the ambient
medium \citep{elmegreen02}.

\subsection{Solutions with the FUV radiation field: what they mean about the H$_2$ layer}
\label{secondsol}

The red dashed curve in Fig. \ref{hunter_q_SF_radial} uses the FUV radiation profile
and the result is a nearly constant $\epsilon_{\rm ff}$ with radius. The constant
$n(H_2)/n(HI)$ in this case follows from equation (\ref{meanmole}) with $\Phi_{\rm
FUV}\approx j_{\rm FUV}/\kappa_{\rm FUV}$ if we use the observation in the bottom
left panel of Figure \ref{hunter_q_sr_radials} that the FUV flux density
$\propto\exp(-1.5R/R_{\rm D})$ and continue to use $H_{\rm
star}\propto\exp(0.5R/R_{\rm D})$, which gives $j_{\rm FUV}\propto\exp(-2R/R_{\rm
D})$. Taking $\kappa_{\rm FUV}$ proportional to $n(HI)\propto\exp(-R/R_{\rm D})$ as
before then gives $\Phi_{\rm FUV}\propto\exp(-R/R_{\rm D})$. Thus $n(HI)/\Phi$ in
equation (\ref{meanmole}) is independent of radius.

This result suggests an interesting possibility if we again match the observed
radial profiles, but now use $j_{\rm FUV}$ for the evaluation of the molecular
fraction. Considering only the H$_2$ layer as a determinant of the SFR, the areal
SFR is
\begin{equation}
\Sigma_{\rm SFR}=\epsilon_0\Sigma_{\rm gas}(H_2)/\tau_{\rm ff}(H_2),
\label{sfrmole}
\end{equation}
where $\Sigma_{\rm gas}(H_2)$ is the column density of H$_2$, $\tau_{\rm
ff}(H_2)=(32 G \rho[H_2]/3\pi)^{-0.5}$ is the free-fall time at the average H$_2$
density, and $\epsilon_0$ is a constant. In this interpretation, the molecular gas
is made explicitly relevant for star formation, rather than through a molecular
fraction proportional to $\epsilon_{\rm ff}$ as above.

Now we have to imagine what the H$_2$ clouds are like in order to derive
$n(H_2)/n(HI)$.  They should still be mostly diffuse except for occasional
self-gravitating cores where stars form, because the interstellar gas density is low
on average. Presumably the cores contain CO molecules. Equation (\ref{formation})
should apply to most regions, but now we might consider a non-negligible molecular
fraction, which, after solving, becomes
\begin{equation}
{{n(H_2)}\over{n(HI)}} \sim {1\over{\Phi\sigma_{\rm dis}/n(HI)D_{\rm DTG}R_{\rm
form}-2}}.
\label{largeh2}
\end{equation}
This equation was obtained by setting $n({\rm dust})=\left(n(HI)+2
n(H_2)\right)D_{\rm DTG}$. In the previous solution, $\Phi\sigma_{\rm
dis}/n(HI)D_{\rm DTG}R_{\rm form}>>2$ so $n(H_2)/n(HI)<<1$. Now with $\Phi_{\rm
FUV}/n(HI)$ approximately constant with radius, $n(H_2)/n(HI)$ is still constant,
and equation (\ref{largeh2}) allows it to be unity or larger.

A constant or high molecular fraction seems out of place for dIrr galaxies, where CO
has barely been detected yet. However, there is growing evidence that $H_2$ might be
much more abundant than we have been expecting. \cite{leroy08} hinted at this given
the high SFR per unit HI in dIrrs, and there is also direct evidence for substantial
far-infrared emission and dust, which implies a high gas content, even when the HI
emission is low, as referenced in Section \ref{mdc} \citep[see review
in][]{bolatto13}.

The next step is to consider what the profile of $\Sigma_{\rm gas}(H_2)$ might be,
as that involves the scale height of the molecular layer in addition to the 3D
density, $n(H_2)$, which comes independently from equation (\ref{formation}). Here
we call upon an important difference between cool HI and H$_2$: the atomic gas has a
warm thermal phase that maintains a high (thermal) velocity dispersion and a large
scale height. Cool HI clouds forming by thermal instabilities in the warm HI layer
can appear at any height where the local pressure supports two thermal phases. In
this sense, the scale height of cool diffuse HI clouds is not directly the result of
some driven turbulent speed, but more the result of the thermal speed of the warm
HI, from which the clouds condense and fall down.  Such falling could be a source of
turbulence in the cool HI medium. The H$_2$ layer is different because it has no
pervasive and warm thermal phase from which cool H$_2$ clouds condense. H$_2$ forms
where the cool HI is shielded from starlight. The molecular clouds can be confined
to the midplane for example, with arbitrarily low turbulent speed and
correspondingly thin line-of-sight depth. The thickness of the H$_2$ layer could be
the result of opacity combined with turbulence and direct stirring, since the
thermal speed of H$_2$ is low.

With this in mind, we consider that the H$_2$ layer thickness does not flare as much
as the HI layer, so that $\Sigma_{\rm gas}(H_2)$ scales almost directly with
$n(H_2)$, i.e., as $\exp(-R/R_{\rm D})$. Recall, that this is the radial dependence
of $n(HI)$ from Figure \ref{hunter_q_sr_radials} and $n(H_2)/n(HI)\sim$ constant in
our model with FUV irradiance. A thin sub-layer of H$_2$ in the midst of a flaring
layer of HI does not violate our assumptions in deriving the molecular fraction,
since only the average midplane density of n(HI) was used. Also, the opacity in our
galaxies is so low, with optical depths larger than any of these thicknesses, that
the vertical distribution of the gas does not much affect the radiation field
$\Phi$.

What it means physically for $\Sigma_{\rm gas}(H_2)$ to be proportional $n(H_2)$ is
that there is a constant number of H$_2$ clouds per line of sight through a galaxy,
whereas with $\Sigma_{\rm gas}(HI)$ proportional to $n(HI)H_{\rm gas}$, the number
of standard HI clouds per line of sight increases with radius as H$_{\rm gas}$
increases. The H$_2$ clouds can be at any height above the plane, i.e., they do not
have to be exactly in the midplane, but they need a total thickness that is
approximately independent of radius. Stars forming in the H$_2$ clouds should have
the same range of heights as the clouds, possibly filling the stellar scale height
$H_{\rm star}$ after 100 Myr, which is the stellar age range observed in the FUV
band.

This model gives the radial dependence of the FUV and SFR, $\exp(-1.5R/R_{\rm D})$,
as observed in Figure \ref{hunter_q_sr_radials}, using the molecular form of the
SFR, equation (\ref{sfrmole}), because
\begin{equation}
\Sigma_{\rm gas}(H_2)\propto n(H_2) \propto n(HI) \propto e^{-R/R_{\rm D}}
\end{equation}
and
\begin{equation}
1/\tau_{\rm ff}(H_2)\propto n(H_2)^{0.5} \propto n(HI)^{0.5} \propto e^{-0.5R/R_{\rm D}}.
\end{equation}

This model also gives $\Sigma_{\rm gas}(H_2)/\Sigma_{\rm gas}(HI)\propto
\exp(-0.5R/R_{\rm D})$ as before (this was $\epsilon_{\rm ff}$ in the model of
Section \ref{radial}), but now it is from $H(H_2)/H_{\rm gas}(HI)$ with constant
H$_2$ layer thickness.  Before, $n(H_2)/n(HI)$ varied with radius this way, with
$H(H_2)\propto H(HI)$.  Thus in both cases, the integrated molecular fraction
decreases with radius with twice the V-band scale length, but when the FUV band
determines the 3D molecular fraction, the local molecular fraction in the region of
the H$_2$ clouds is about constant with radius. Also in both cases, the density used
for the dynamical time, $n(HI)$ in Section \ref{radtrans} and $n(H_2)$ in the
present section, is proportional to total interstellar pressure as determined from
the column density.

\subsection{Avoiding a sharp transition in the molecular fraction of the ISM}
\label{sharp}

The inferred variations of FUV volume emissivity, $j_{\rm FUV}\propto\exp(-2R/R_{\rm
D})$, and midplane gas density, $n(HI)\propto\exp(-R/R_{\rm D})$, produce a ratio
$j_{\rm FUV}/n(HI)^2$ that is constant with radius. If we continue to identify the
local radiation field, $\Phi$, with $j_{\rm FUV}/\kappa_{\rm FUV}$ and if
$\kappa_{\rm FUV}\propto n(HI)$, then $\Phi/n(HI)$ is constant. This is the primary
assumption made by \cite{krum09,krumholz09b} in their derivation of molecular
self-shielding in clouds. With this assumption, the ISM can maintain two thermal
phases and cool HI clouds are possible. This is an important aspect of our model
also because cool HI clouds are the source of molecules and eventual star formation
even in a diffuse molecular medium.

If we consider molecule formation on a cloud-by-cloud basis, which means that clouds
are either highly molecular beyond some column density threshold or highly atomic
below this threshold, then we would obtain a certain variation of average molecular
fraction with galactocentric radius that may be compared with the observations. In
\cite{krumholz09b}, such an assumption produces a sharp radial transition in the
overall molecular fraction of the ISM, and this sharpness is offered as an
explanation for the break in the slope of the star formation -- column density
relation between the inner and outer disks (see Section \ref{intro}). A sharp
transition is inconsistent with the smooth variations in FUV and gas properties that
are observed in our dwarf galaxy sample here.

To be more specific, we consider two examples. We first model the formation of
molecules according to equation (\ref{formation}), which may be re-written in terms
of a threshold column density for individual clouds to shield themselves and form
H$_2$. Writing this threshold as $\Sigma_{\rm thresh}= (n(HI)+2n(H_2))L$ for cloud
depth $L$, we obtain
\begin{equation}
\Sigma_{\rm thresh}\sim {{\Phi}\over{n(HI)R_{\rm form}D_{\rm DTG}}},
\label{thresh}
\end{equation}
where $n(H_2)\sigma_{\rm dis}L\sim1$ at the threshold because this is the optical
depth for self-shielding.  If $\Phi/n(HI)\sim$constant with radius in the FUV case,
then the shielding threshold is also about constant.  The column density for
shielding is about constant in the model by \cite{krumholz09b} too, as given by
their equations (12) or (49) and their Figure 4 at high $\Sigma_{\rm obs}$, because
they also have a constant ratio of photon to gas density.

To see the implications of the shielding layer on a population of clouds, we set the
radial profile of the molecular fraction in a galaxy equal to the radial profile of
the mass fraction of all of the shielded parts of clouds. Clouds with column
densities greater than $\Sigma_{\rm thresh}$ are assumed to have the excess column
density in molecular form and the rest atomic, while clouds with $\Sigma<\Sigma_{\rm
thresh}$ are assumed to be totally atomic. This means that as the pdf for cloud
column density (Fig. \ref{figure9_china2012}) shifts toward lower mean values with
increasing radius, the integral above $\Sigma_{\rm thresh}$ of the pdf multiplied by
$\Sigma-\Sigma_{\rm thresh}$, which is the molecular part, decreases also. The
log-normal form of this pdf causes the integral to drop abruptly at some radius.

The molecular fraction in this threshold model is given by
\begin{equation}
{{n(H_2)}\over{n_{\rm total}}}={{\int_{\Sigma_{\rm thresh}}^\infty \left(\Sigma -\Sigma_{\rm thresh}\right)
P(\log\Sigma)d\log\Sigma}\over{\int_0^\infty \Sigma P(\log\Sigma)d\log\Sigma}}
\end{equation}
where
\begin{equation}
P(\log\Sigma)=P_0e^{-0.5\left(\log\Sigma/\Sigma_{\rm peak}\right)^2/S^2}.
\end{equation}
The dispersion $S$ is usually written in terms of an effective average Mach number
\citep{pnj97}, $M$,
\begin{equation}
S=(\ln(1+0.25M^2))^{0.5},
\end{equation}
and the column density at the peak is proportional to the average column density of
clouds,
\begin{equation}
\Sigma_{\rm peak}=\Sigma_{\rm ave}e^{-0.5S^2}.
\end{equation}
The radial variation of $\Sigma_{\rm ave}$ is not known as this is for individual
clouds rather than for the whole ISM, which is what we measure directly. We assume
in two models that it varies either as the observed HI column density, which is
approximately $\propto e^{-0.5R/R_{\rm D}}$ in the star-forming region (Sect.
\ref{radial} and Table 2), or as the inferred HI midplane density, which is
approximately $\propto e^{-R/R_{\rm D}}$. What enters then is the dimensionless
ratio of the average central column density to the threshold column density for
$H_2$, $A=\Sigma_{\rm gas}(R=0)/\Sigma_{\rm thresh}$.

Figure \ref{hunter_q_kappa} shows the molecular fraction versus radius for one case
where $\Sigma_{\rm ave}$ follows the observed HI column density (dashed lines) and
another case where $\Sigma_{\rm ave}$ follows the inferred HI midplane density
(solid lines). Blue curves assume $A=10$ so the central region is highly molecular,
and red curves assume $A=2$. Both use $M=4$ to get some breadth in the pdf.
According to the discussion in Section \ref{radial}, the deviation between the
observed star formation rate and the dynamical rate, $\Sigma_{\rm gas}/t_{\rm ff}$,
varies approximately exponentially with radius, and we interpreted this as a
possible variation in the fraction of the ISM in the form of cold molecular clouds
-- molecular because this additional variation is not seen in the HI profiles, and
cold because they are eventually involved with star formation, perhaps after 3D
processes with self-gravity. Such a variation would be a straight line in Figure
\ref{hunter_q_kappa}, and not a steeply falling curve. The falling curve means that
the H$_2$ fraction has a sharp cutoff at some radius where the average column
density of a cloud is comparable to the threshold column density in the local
radiation field.

A second model with an H$_2$ cutoff is based on the theory in
\cite{krum09,krumholz09b} which considers the molecular content of individual clouds
in more detail. That model calculates the HI boundary layer of spherical clouds
using H$_2$ self-absorption and dust absorption, and it uses this layer to determine
the molecular fraction in the cloud given reasonable assumptions about the ratio of
H$_2$ and HI densities. The model also assumes a constant ratio of radiation density
to average gas density in order to maintain a 2-phase ISM.

In \cite{krumholz09b}, the molecular fraction in a cloud approaches unity when their
variable $s$ greatly exceeds 1, where $s=\Sigma_{\rm comp}Z/\psi$ for cloud column
density $\Sigma_{\rm comp}$ in units of $M_\odot$ pc$^{-2}$, $Z$ is the metallicity
in solar units, and $\psi$ is a variable of order unity that depends on the ratio of
radiation density to gas density. For their fiducial model with $Z\sim1/8$ as for
our galaxies, their equation (7) gives $\chi=1.88$ for $\phi_{\rm CNM}=3$ and
standard absorption and formation coefficients for H$_2$, and their equation (10)
then gives $\psi=1.08$. Thus $s=0.12\Sigma_{\rm comp}$. Now looking at their
equation (39) for the quantity $R_{\rm H_2}=n(H_2)/n(HI)$, and writing
\begin{equation}
{{n(H_2)}\over{n_{\rm total}}}={{R_{\rm H_2}}\over{1+R_{\rm H_2}}}
\end{equation}
we see that this molecular fraction approaches unity for $s>>11$, which corresponds
in our case to $\Sigma_{\rm comp}>>100\;M_\odot$ pc$^{-2}$. At $s<<8.4$, the
molecular fraction goes to zero as $(s/12.1)^3$, which is $(\Sigma_{\rm
comp}/100)^3$. At low molecular fraction, this expression should be considered an
upper limit.  The result is a rapid drop in molecular fraction with the third power
of the decreasing gas column density.

A similar rapid drop in molecular fraction is obtained from equation (2) in
\cite{krum09}. We evaluate that equation again with metallicity $Z=1/8$, which gives
$\chi=1.89$, $s=151/\Sigma_{\rm comp}$, $\delta\sim0.21$ for large $s$, and
molecular fraction approaching zero for low $\Sigma_{\rm comp}/151$ as $(\Sigma_{\rm
comp}/129)^5$.

These rapid drops in molecular fraction for the pdf-threshold model and for the
\cite{krumholz09b} and \cite{krum09} models may explain the transition between the
inner molecule-rich and outer molecule-poor regions of spiral galaxies, which seems
to be relatively sharp in the observations by \cite{bigiel08} and others. Different
processes seem to be at work for dIrr galaxies and perhaps also for the far-outer
regions of spirals where the variations in SFR appear to be smoother. This is the
primary reason for the diffuse cloud model in Section \ref{mdc}, which considers
average H$_2$ formation and destruction rates in large volumes rather than threshold
shielding in individual clouds.

\section{Searching for the edges of star-forming disks}
\label{search}

Dwarf irregulars probe star formation at extremely low surface densities, where 2D
processes like spiral waves are inactive (Sect. \ref{two-fluid}) and the gas
consumption time can exceed a Hubble time. In Figure \ref{hunter_q_sr_radials} and
Table 2, our sample is traced out to where the gas surface density is
$\sim1\;M_\odot$ pc$^{-2}$, the stellar surface density is $\sim0.1\;M_\odot$
pc$^{-2}$, and the areal SFR is $\sim10^{-5}\;M_\odot$ pc$^{-2}$ Myr$^{-1}$.  Note
that this SFR multiplied by the Hubble time equals the stellar surface density, so
the observed rate is about equal to the average rate over the life of the galaxy.

These limits were also found in a previous study \citep{hunter11} of four dIrr
galaxies and a BCD galaxy using deep $V$ and $B$ band observations along with FUV
from GALEX and HI from LITTLE THINGS \citep{hunter12} and THINGS \citep{walter08}.
We traced the disks down to about $29.5$ mag arcsec$^{-2}$ in V band with B-V$\sim
0.3–-0.4$, giving a mass-to-light ratio of 1.07 for a constant SFR with a Salpeter
(1955) IMF \citep{bell01}. Then the stellar density is $\sim0.06\;M_\odot$ pc$^{-2}$
($0.03$ for a Chabrier IMF). At the last traceable point of FUV or H$\alpha$, the
gas surface density was $\sim1\;M_\odot$ pc$^{-2}$ and the star formation rate
$\sim10^{-5}\;M_\odot$ pc$^{-2}$ Myr$^{-1}$, as for the present sample.  The gas
surface density continued down to $0.1\;M_\odot$ pc$^{-2}$ in DDO 53 and DDO 133,
which are in the present sample too.

The same limits apparently apply to spiral galaxies. In M33, \cite{grossi11} found
that star formation extends for about 10 inner scale lengths, to a radius of 60
arcmin where $\Sigma_{\rm HI}\sim1\;M_\odot$ pc$^{-2}$ and $\Sigma_{\rm
star}\sim0.1\;M_\odot$ pc$^{-2}$.  Both gas and stars go further than this although
the stars have a shallower profile in the outer region. In NGC 7793
\citep{radburn12}, star formation ends before the gas and stars, at $\Sigma_{\rm
gas}\sim1\;M_\odot$ pc$^{-2}$, $\Sigma_{\rm star}\sim0.1\;M_\odot$ pc$^{-2}$, and
$\Sigma_{\rm SFR}\sim10^{-4.5}\;M_\odot$ pc$^{-2}$ Myr$^{-1}$. In NGC 2403
\citep{barker12}, star formation goes for 8 exponential scale lengths down to a
V-band surface brightness of $\sim29.5$ mag arcsec$^{-2}$ and old stars continue
further with a shallower slope, so this is about the same limit. In \cite{hunter13},
NGC 801 and UGC 2885 have observed star formation to the same limits in $\Sigma_{\rm
gas}$, $\Sigma_{\rm star}$, and $\sigma_{\rm SFR}$.

Gas mass dominates stellar mass in these far outer regions, so the disk total gas
pressure, scale height, and midplane density follow from the equations for a gaseous
disk with a velocity dispersion $\sigma_{\rm gas}$. When $\Sigma_{\rm
gas}=1\;M_\odot$ pc$^{-2}$, the pressure is $(\pi/2)G\Sigma_{\rm gas}^2= 34 k_{\rm
B}$ for Boltzmann constant $k_{\rm B}$, the scale height is $\sigma_{\rm gas}^2/\pi
G \Sigma_{\rm gas}=4.7$ kpc, and the midplane mass density is $0.5\pi G (\Sigma_{\rm
gas}/\sigma_{\rm gas})^2= 7.22\times10^{-27}$ g cm$^{-3}$, which corresponds to
$n(HI)=0.0033$ cm$^{-3}$. Here we have assumed a gaseous velocity dispersion
$\sigma_{\rm gas}=8$ km s$^{-1}$ for the far outer parts from Table 2.  Also from
these numbers, $\tau_{\rm ff}=780$ Myr. If $\Sigma_{\rm SFR}=\epsilon\Sigma_{\rm
gas}/\tau_{\rm ff}$, then $\epsilon=7.8\times10^{-3}$ when $\Sigma_{\rm
SFR}=10^{-5}\;M_\odot$ pc$^{-2}$ Myr$^{-1}$. This value of $\epsilon$ is consistent
with the values found at large radii in the lower right panel of Figure
\ref{hunter_q_SF_radial}.

Stars and HI gas continue further than star formation in most of the galaxies
mentioned above, and they do so for many in the present sample as well. The radial
profiles in Figure \ref{hunter_q_sr_radials} are plotted only out to the edge of the
FUV light where the SFR can be measured. Figure \ref{hunter_q_radial_Nsnt} plots the
HI and derived data out as far as the observations go.  The 3D density is determined
from $\Sigma_{\rm gas}$ assuming that gas dominates stars, as above, and using the
observed velocity dispersion at the appropriate radius. The free fall time comes
from the density. In some cases, the full HI radial profiles go out nearly twice as
far as the star forming parts shown in Figure \ref{hunter_q_sr_radials}, and the
slopes change in the outer parts.  One of the red lines in each panel is the slope
tabulated in Table 2, which is from the radial range where FUV is detected, while
another red line suggests a different slope, more appropriate to the outer regions.

The smooth exponential profiles and large $\tau_{\rm ff}$ for the outer regions
shown in Figure \ref{hunter_q_radial_Nsnt} call into question some of the modern
ideas about cosmic accretion. Cosmic accretion seems required to explain the
continuous presence of star formation in galaxies of moderate to low mass when the
gas consumption time is only $\sim0.2$ times the current Hubble time
\citep{dekel13}. For $\sim2$\% efficiency, the consumption time is $50\tau_{\rm
ff}$. This replenishment scenario applies mostly to the inner parts of spirals.
However, when the consumption time is comparable to or longer than the Hubble time,
accretion is not needed for the same reason, and may even result in too much gas
collecting in the outer parts. In Figure \ref{hunter_q_radial_Nsnt}, the HI profiles
do not show an excess when $\tau_{\rm ff}$ gets large, they continue to decline with
about the V-band scale length in most cases (which is even steeper than for the
inner regions). Three galaxies in the upper left panel have more slowly declining
$\Sigma_{\rm gas}$ and $n(HI)$; these are DDO 75 (Sextans A), DDO 154 (NGC 4789A),
and Haro 29.


As mentioned above, stellar disks often continue further out than the known star
formation, in which case the disks cannot be made there but have to move there.
Outer disks also have smooth exponential surface density profiles like inner disks.
Bars \citep{debattista06} and spiral waves \citep{barrier14} could be responsible
for the exponential profiles in the main disks of spiral galaxies and barred dIrr
galaxies, and these structures might even scatter stars to the far outer regions
where star formation stops \citep{roskar08}. However, in non-barred dIrr galaxies,
the only irregular structures that seem capable of scattering stars are gas clumps,
which might also make exponentials \citep{elmegreen13c}.

Our observation that star formation follows the dynamical time at the midplane
density implies that 3D self-gravity in the gas is still important even without
spirals. This 3D gravity should make dense clouds throughout the disks in our
sample. Moreover, the dominance of gas in the outer parts (the midplane gas-to-star
density ratio is $\sim10$ according to Table 2) implies that such clouds are the
main scattering sites for ambient stars.  Thus the outer disk profiles of old stars
could be the result of stellar scattering off of massive conglomerations of diffuse
clouds.

\section{Conclusions}
\label{conc}

The star formation properties of 20 dIrr galaxies have been examined in relation to
the average radial profiles of gas and stellar mass, FUV intensity, velocity
dispersion, and rotation rate. The disks are found to be relatively thick and stable
against 2D self-gravitational processes, which probably explains why there are no
spiral waves, but which also suggests that star formation is fundamentally a 3D
process. The star formation rate agrees fairly well with the conventional
expression, $\epsilon_{\rm ff}\Sigma_{\rm gas}/\tau_{\rm ff}$, also giving the usual
efficiency, $\epsilon_{\rm ff}$, of one per cent, when the free fall time,
$\tau_{\rm ff}$, uses the 3D midplane density evaluated from the HI column density
divided by twice the HI scale height. The resulting correlation is fairly tight and
the same for all galaxies, suggesting that ambient self-gravity in 3D is important
for regulating the SFR. We find a radial variation in $\epsilon_{\rm ff}$ that has
an exponential form with a scale length equal to about twice the V-band scale
length. The origin of this $\epsilon_{\rm ff}$ variation is not obvious as it is
likely to come from a combination of effects including radial variations in the
molecular fraction and in the fraction of the gas that is cold and dense enough to
collapse into stars.  A simple model considering only the molecular fraction in a
diffuse interstellar medium from a balance between H$_2$ formation at the rms
density and H$_2$ destruction at the mean density and radiation field, gets about
the observed trend when the atomic and molecular layer thicknesses increase with
radius in the same way.

Another model using the radial profile of FUV radiation to determine the trend in
molecular fraction gives the observed SFR profile too, but only when the molecular
layer does not flare like the HI layer. This seems to be a reasonable proposition
that might be tested in edge-on dIrr galaxies. In this second model, the SFR is
proportional to the ratio of the molecular surface density to the molecular
dynamical time with a constant efficiency.  In the absence of a molecular flare,
this ratio is the same as the 1.5 power of the molecular surface density. This
scaling with molecules is different from that in the inner parts of spiral galaxies
where the SFR seems to scale with the 1st power of the molecular density (see
references in Section 1), but there are other differences between these two regions
also, such as the lack of a gaseous flare, a closer correspondence between H$_2$ and
the more easily observed CO, and a higher density, opacity and degree of
self-gravity for the inner regions of spirals.

Other models that consider a threshold for the presence of H$_2$ in clouds, as might
arise from a shielding layer around the molecular part of a massive cloud, imply
sharp declines in the radial profiles of the molecular fraction that are not
inferred from the smooth profiles of everything else observed here.  Perhaps these
threshold models explain the difference between the inner and outer disk properties
in spiral galaxy, but they do not appear to apply to dIrr disks alone.

The radial profiles of the gas surface densities are generally shallower than the
radial profiles of the V-band light, which are shallower still than the radial
profiles of the FUV light. All have approximately exponential shapes, and the scale
lengths center around factors of 2, 1, and 2/3, respectively, times the V-band scale
length, $R_{\rm D}$. The radial profiles of the gaseous and stellar scale heights
are exponential also, and similar to each other with a scale length of twice $R_{\rm
D}$.  As a result, the midplane densities for stars and gas have different profiles
than the projected densities. For example, volume emissivities for V and FUV bands
drop faster than their surface brightnesses, and the radiation fields in these
bands, given approximately by the volume emissivity divided by the opacity, drop
slower than the surface brightnesses. Radiation fields are stronger in the outer
parts of galaxies than the surface brightness profiles suggest.

The conditions in our galaxies could be representative of the conditions in the
far-outer parts of spiral galaxies, where gas also tends to dominate stars and
thickness effects become important for stabilization. Spiral galaxies still have
spiral waves in the outer parts, however, while dIrrs do not, so there could be a
difference in how GMCs are assembled in the two cases.

\clearpage

\begin{center}
\begin{deluxetable}{lcccc}
\tablenum{1} \tablecolumns{5} \tablewidth{315pt} \tablecaption{The Galaxy Sample
\label{tab-sample} } \tablehead{
\colhead{} & \colhead{D} & \colhead{M$_V$} & \colhead{$R_D$\tablenotemark{a}} & \colhead{log SFR$_D^{FUV}$\tablenotemark{b}} \\
\colhead{Galaxy} & \colhead{(Mpc)} & \colhead{(mag)} & \colhead{(kpc)} &
\colhead{(M$_\odot$ yr$^{-1}$ kpc$^{-2}$)} } \startdata
CVnIdwA   &  3.6 & -12.4 & $0.57 \pm 0.12$  & $-2.48 \pm 0.01$ \\
DDO 101   &  6.4 & -15.0 & $0.94 \pm 0.03$  & $-2.81 \pm 0.01$ \\
DDO 126   &  4.9 & -14.9 & $0.87 \pm 0.03$  & $-2.10 \pm 0.01$ \\
DDO 133   &  3.5 & -14.8 & $1.24 \pm 0.09$  & $-2.62 \pm 0.01$ \\
DDO 154   &  3.7 & -14.2 & $0.59 \pm 0.03$ & $-1.93 \pm 0.01$ \\
DDO 168   &  4.3 & -15.7 & $0.82 \pm 0.01$ &$-2.04 \pm 0.01$ \\
DDO 210   &  0.9 & -10.9 & $0.17 \pm 0.01$ & $-2.71 \pm 0.06$ \\
DDO 216   & 1.1  & -13.7 & $0.54 \pm 0.01$ & $-3.21 \pm 0.01$ \\
DDO 50     &  3.4  & -16.6 & $1.10 \pm 0.05$ & $-1.55 \pm 0.01$ \\
DDO 52     & 10.3 & -15.4 & $1.30 \pm 0.13$ & $-2.43 \pm 0.01$ \\
DDO 53     &  3.6  & -13.8 & $0.72 \pm 0.06$ &$-2.41 \pm 0.01$ \\
DDO 70     &  1.3  & -14.1 & $0.48 \pm 0.01$ & $-2.16 \pm 0.00$ \\
DDO 75     &  1.3  & -13.9 & $0.22 \pm 0.01$ & $-1.07 \pm 0.01$ \\
DDO 87     &  7.7  & -15.0 & $1.31 \pm 0.12$ & $-1.00 \pm 0.01$ \\
Haro 29     &  5.8  & -14.6 & $0.29 \pm 0.01$ & $-1.07 \pm 0.01$ \\
IC 1613     &  0.7  & -14.6 & $0.58 \pm 0.02$ & $-1.99 \pm 0.01$ \\
NGC 2366 &  3.4 & -16.8 & $1.36 \pm 0.04$ & $-1.66 \pm 0.01$ \\
NGC 3738 &  4.9 & -17.1 & $0.78 \pm 0.01$ & $-1.53 \pm 0.01$ \\
UGC 8508 &  2.6 & -13.6 & $0.27 \pm 0.01$ & \nodata                  \\
WLM           & 1.0  & -14.4 & $0.57 \pm 0.03$ & $-2.05 \pm 0.01$ \\
\enddata
\tablenotetext{a}{$R_D$ is the disk scale length measured from $V$-band images. From
Hunter \& Elmegreen (2006).} \tablenotetext{b}{ SFR$_D^{FUV}$ is the SFR determined
from {\it GALEX} FUV fluxes (Hunter et al.\ 2010), with an update of the  {\it
GALEX} FUV photometry to the GR4/GR5 pipeline reduction presented in Zhang et al.\
(2012).}
\end{deluxetable}
\end{center}

\begin{deluxetable}{llcc}
\tablenum{2} \tablecolumns{4} \tablewidth{430pt}
\tablecaption{Radial Trends \label{tab-trends} }
\tablehead{
\colhead{V\tablenotemark{a}} &
\colhead{$\ln \;V(0)$} &
\colhead{$V(0)$ -- $V(4R/R_{\rm D})$} &
\colhead{Slope}}
\startdata
$\Sigma_{\rm gas}$  & $2.28\pm0.69$ & 9.77 -- 1.589$\;M_\odot$ pc$^{-2}$& $-0.45\pm0.32$\\
$\Sigma_{\rm star}$ & $1.58\pm1.26$ & 4.87 -- 0.134$\;M_\odot$ pc$^{-2}$& $-0.90\pm0.36$\\
$n_{\rm gas}$       & $-0.83\pm0.91$ &0.44 -- 0.0116 cm$^{-3}$& $-0.91\pm0.54$\\
$n_{\rm star}$      & $-1.23\pm1.52$ &0.29 --   0.0011 cm$^{-3}$& $-1.40\pm0.49$\\
$\sigma_{\rm gas}$ & $ 2.47\pm0.34$ & 11.8 --  8.0 km s$^{-1}$ & $-0.10\pm0.10$\\
$H_{\rm gas}$  & $ 48.42\pm0.59$ &0.35 -- 2.13 kpc& $ 0.45\pm0.28$\\
$H_{\rm star}$ & $ 48.12\pm0.71$ &0.26 -- 1.88 kpc& $ 0.50\pm0.25$\\
$n_{\rm star}/n_{\rm gas}$ & $-0.40\pm1.26$ & 0.67 -- 0.0954& $     -0.49\pm 0.51$\\
$\Sigma_{\rm gas}/\tau_{\rm ff}$ & $ -1.94\pm1.08$ &0.144 --  0.0038 $M_\odot$ pc$^{-2}$ Myr$^{-1}$ & $-0.91\pm0.58$\\
$\Sigma_{\rm SFR}$ & $-5.75\pm1.25$ & $0.0032$ --  $1.03\times10^{-5}$ $M_\odot$ pc$^{-2}$ Myr$^{-1}$ & $-1.43\pm0.58$\\
$\tau_{\rm ff}$ & $35.30\pm34.43$ &0.068 --  0.42 Gyr & $ 0.45\pm0.27$
\enddata
\tablenotetext{a}{$V$ represents the ``variable'' given in the first column; the second column
is the natural log of this variable in the galaxy center, the 3rd column is the range out to
4 scale lengths, and the 4th column is the slope on a $\ln$-linear plot. The exponential V-band
light profile would have a slope of $-1$.}
\end{deluxetable}

\clearpage

\clearpage

\clearpage

\clearpage

\clearpage

\clearpage

\clearpage
\begin{figure}\epsscale{.4}
\plotone{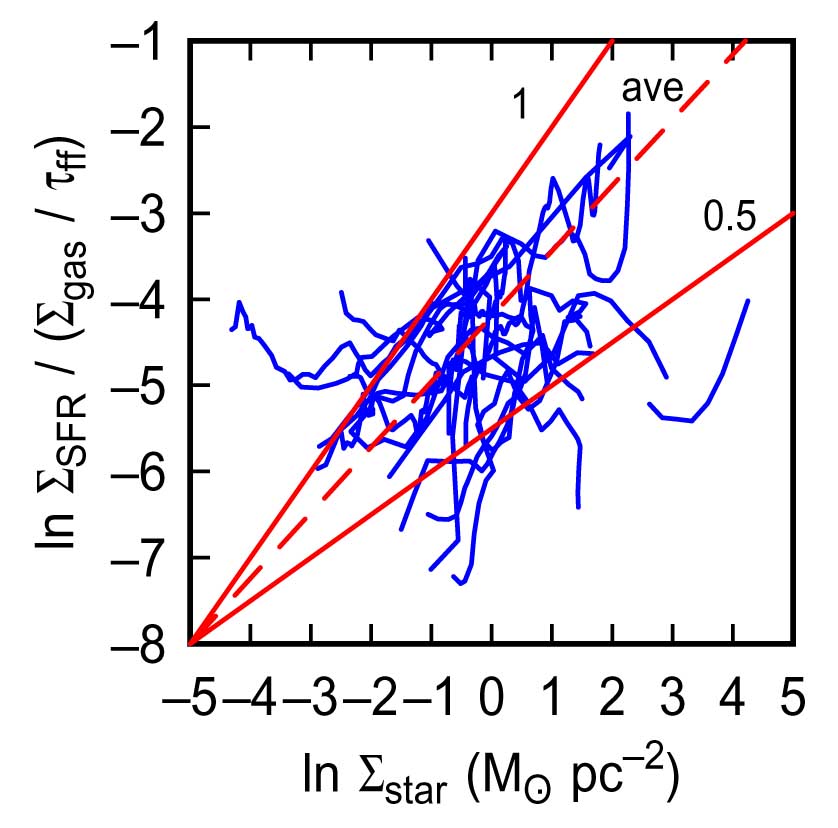}
\caption{Star formation efficiency versus the stellar mass surface density,
with natural logs.
} \label{hunter_q_shi}\end{figure}

\clearpage
\begin{figure}\epsscale{.9}
\plotone{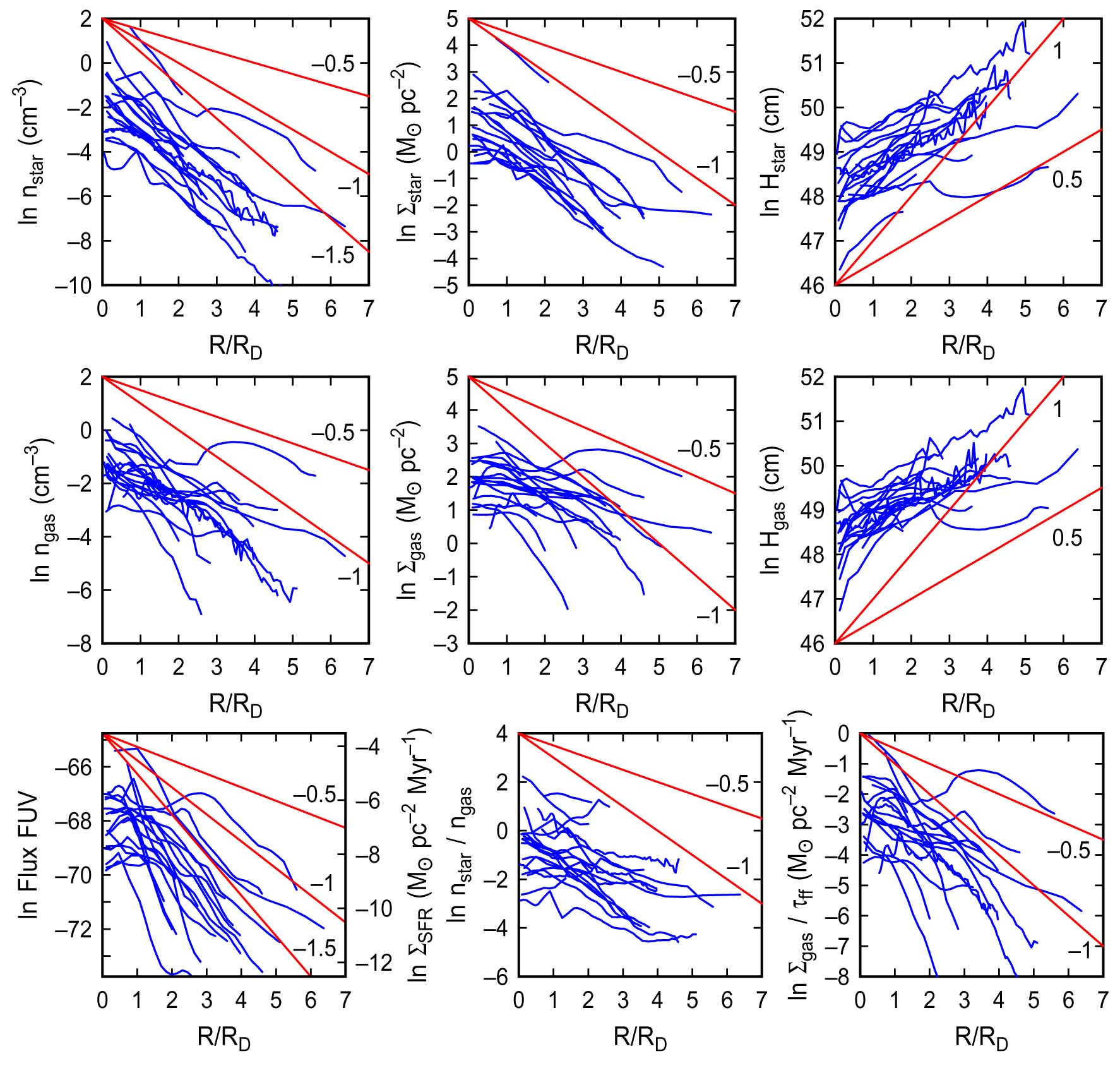}
\caption{Radial profiles of various quantities plotted with natural
log on the ordinate to facilitate comparisons with the exponential disk profile.
The straight lines have slopes in increments of $0.5$.
} \label{hunter_q_sr_radials}\end{figure}

\clearpage
\begin{figure}\epsscale{.6}
\plotone{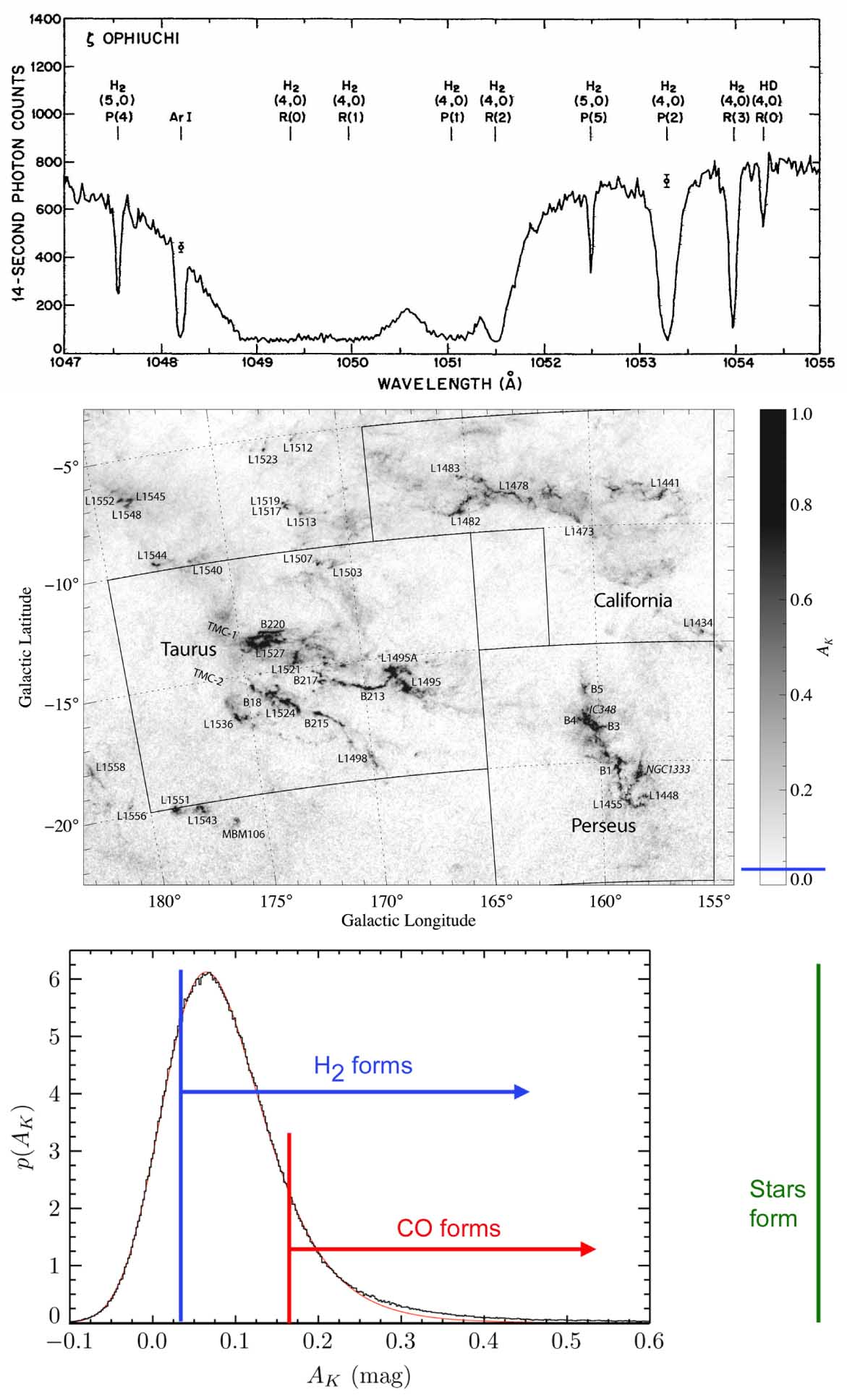}
\caption{(top) Spectrum of molecular hydrogen absorption on the line of
sight to $\zeta$Oph showing nearly complete absorption from
a standard-size diffuse cloud. (middle) The Perseus and Taurus regions
studied by Lombardi et al. (2010) with the threshold extinction for H$_2$ formation
indicated by a horizontal bar on the gray-scale calibration.
Most of the field of view has an extinction exceeding the threshold. (bottom)
Distribution function for column density in the Perseus-Taurus region, from
Lombardi et al. (2010) showing the H$_2$, CO, and star formation thresholds
discussed in the text. A large fraction of the diffuse interstellar
medium in the solar neighborhood is H$_2$.} \label{figure9_china2012}\end{figure}

\clearpage
\begin{figure}\epsscale{.6}
\plotone{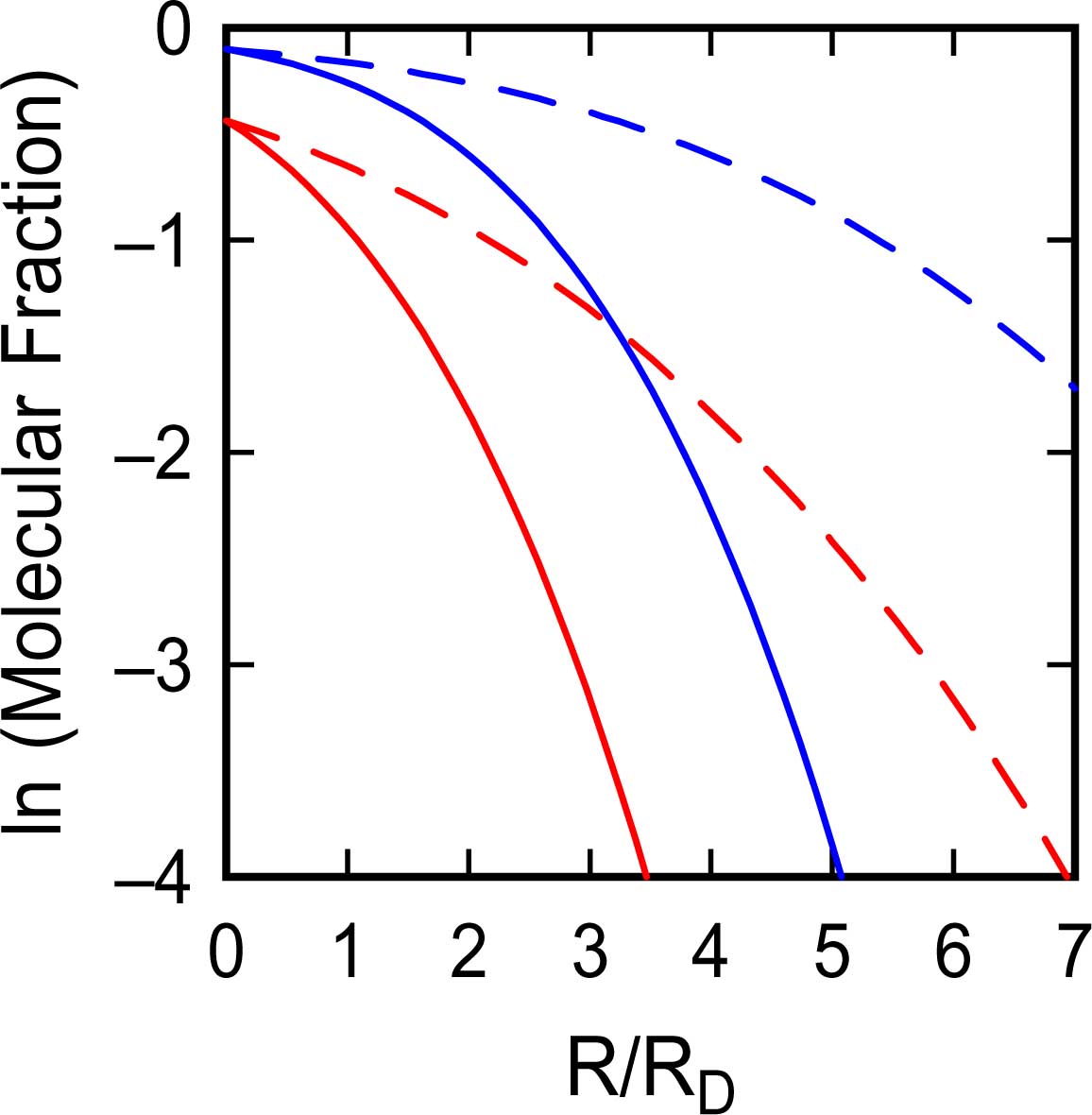}
\caption{Radial profiles of molecular fraction in two threshold models for H$_2$
formation. Blue curves are for a highly molecular ISM in the center, where the
ratio of the mean surface density of gas to the H$_2$ formation threshold equals 10,
and red curves are for a moderate molecular fraction in the center, where this
ratio equals 2. Dashed curves assume the average cloud column density scales with
the total HI surface density and solid curves assume the average cloud column
density scales with the midplane 3D density. } \label{hunter_q_kappa}\end{figure}

\clearpage
\begin{figure}\epsscale{.8}
\plotone{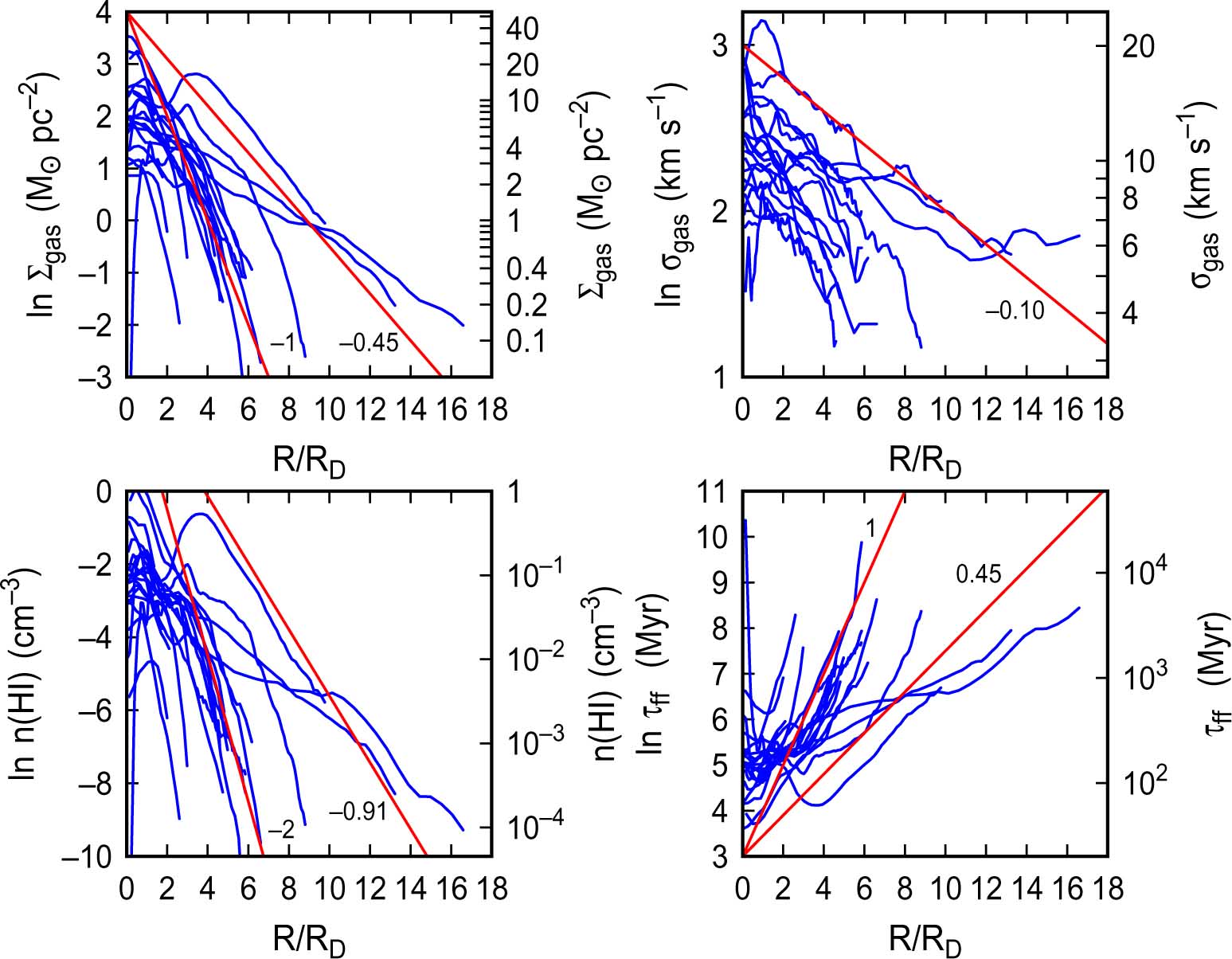}
\caption{Radial profiles for quantities derived from the HI observations: surface
density in the upper left, velocity dispersion in the upper right, 3D midplane
density in the lower left and free fall time in the lower right. The plots are
in log-linear space with the natural logarithms on the left-hand axes for comparison
to the exponential disk profiles discussed elsewhere. The right-hand axes are
plotted in physical units, again on a log scale. The slopes of the red lines are
indicated. One red line in each panel has a slope given by a fit to the inner
regions (Table 2) and another red line, if there is one,
suggests a different slope for the gas in the outer
region. } \label{hunter_q_radial_Nsnt}\end{figure}

\end{document}